\documentclass{iopjournal}

\usepackage{amsmath,amssymb}
\usepackage{subcaption}

\usepackage{color}
\definecolor{mypurple}{rgb}{0.5, 0, 0.85}
\definecolor{Hbetta}{rgb}{0,0.92,1}
\definecolor{myblue}{rgb}{0, 0.2, 0.85}
\definecolor{cadmiumgreen}{rgb}{0.0, 0.42, 0.24}
\definecolor{gold}{rgb}{0.7176, 0.5843, 0.0431}
\definecolor{org}{RGB}{33, 47, 61}

\usepackage[
    style=authoryear,
    maxcitenames=2,
    maxbibnames=10,
    sorting=nyt,
    backend=bibtex
]{biblatex}

\usepackage{aas_macros}

\begin{document}

\articletype{Paper} %	 e.g. Paper, Letter, Topical Review...

\title{Linear Gravitational Wave Memory Through the Window of Core-Collapse Supernovae}

\author{Colter J. Richardson$^{1}$\orcid{0000-0003-1866-7965}, Anthony Mezzacappa$^{1}$\orcid{0000-0001-9816-9741}, Haakon Andresen$^{2}$\orcid{0000-0002-4747-8453}, and Michele Zanolin$^{3}$}

\affil{$^1$Department of Physics and Astronomy, University of Tennessee, Knoxville, TN 37996, USA}

\affil{$^2$The Oskar Klein Centre, Department of Astronomy, AlbaNova, SE-106 91 Stockholm, Sweden}

\affil{$^3$Embry-Riddle Aeronautical University, 3700 Willow Creek Road, Prescott, Arizona 86301, USA}

\email{cricha80@vols.utk.edu}


\begin{abstract}
    Low-frequency gravitational waves ($\lessapprox$ 50 Hz) from core-collapse supernovae are becoming more important for current and future gravitational wave studies. This frequency region is dominated by the global morphology of the explosion and the anisotropic emission of neutrinos from the event. This paper serves as a brief review of both theory and detection (prospects) for gravitational waves in the low-frequency region. We discuss the generation of the linear gravitational wave memory sourced from neutrino emission and show results from an example 15 $M_{\odot}$ Solar metallicity progenitor. We also discuss the detection of the linear gravitational wave memory in current detectors, utilizing a combination of a linear predictive filter and matched templating. Finally we will discuss detection prospects in future detectors such as Cosmic Explorer, Einstein Telescope, the Laser Interferometer Space Antenna, and the Lunar Gravitational-wave Antenna.

\end{abstract}

\section{Introduction}
\label{sec:intro}
    As is well known in this special journal, core-collapse supernovae (CCSNe), the energetic deaths of massive stars, are well poised to emit gravitational radiation that should be detectable in current gravitational wave detectors (see \cite{2016LRR....19....1A}), such as LIGO (see \cite{2015CQGra..32g4001L}), Virgo (see \cite{2015CQGra..32b4001A}), and KAGRA (see \cite{2021PTEP.2021eA101A}) if the event occurs within our galaxy. Due to the rarity of these events in our galaxy, 1-2 per century, we need to prepare for the next CCSN and develop detection algorithms and signal analysis techniques to maximize the scientific gain from the event. There has been significant effort dedicated to the detection and analysis of GWs at intermediate and high-frequencies, see the reviews by \cite{2006RPPh...69..971K, 2022hgwa.bookE..21A, 2025RvMP...97d1002M}. 
    However, recently there has been motivation to study the low-frequency portion of the signal less than approximately 50 Hz, in current and future detectors 
    \cite{1978ApJ...223.1037E, 1978Natur.274..565T, 1996PhRvL..76..352B, 1997A&A...317..140M, 2004ApJ...603..221M, 2009ApJ...697L.133K, 2009ApJ...707.1173M, 2010CQGra..27s4005Y, 2011ApJ...743...30T, 2011ApJ...736..124K, 2012A&A...537A..63M, 2013ApJ...766...43M, 2015PhRvD..92h4040Y, 2019ApJ...876L...9R, 2020MNRAS.494.4665P, 2022PhRvD.105j3008R, 2022MNRAS.510.5535J, 2022PhRvD.106d3020M, 2023PhRvD.107d3008M, 2023MNRAS.522.6070P, 2023ApJ...959...21P, 2023PhRvD.107j3015V, 2024PhRvL.133w1401R, 2024ApJ...975...12C, 2021arXiv210505862M, 2024arXiv240513211G, 2025PhRvD.112l3025R, 2026arXiv260407878S}.
    This article serves to review the progress in the field of low-frequency CCSN gravitational-wave (GW) astronomy and provide predictions for future directions in studying these signals in next-generation detectors. 

    Low-frequency GWs are currently understood to be sourced from two features: 1) The anisotropy of the explosion. 2) The anisotropic emission of neutrinos resulting from the anisotropic explosion or from the anisotropic dynamics before a failed explosion. Our focus herein is the linear GW memory, or permanent deformation of spacetime resulting from a permanent shift in the stress-energy of the system which appears around and below 50 Hz. This portion of the signal has been neglected in current detectors, until recently, due to it encroaching onto the seismic noise dominated 10 Hz wall in the LIGO data (see \cite{2025PhRvD.111f2002C}). The following review will walk through the history of linear GW memory, within the context of CCSNe, the current endeavor for detection strategies, and the future of potential detections.

    Following the realization of general relativity and then wave-solutions to Einstein's equations, \cite{1974SvA....18...17Z} were the first to calculate the gravitational radiation utilizing the linearized form of the Einstein equations and investigating the GW signature from two ``particles'' or dense stars moving from infinity and passing each other in a fly-by. Following this, \cite{1978ApJ...223.1037E} derived an equation for gravitational radiation from the anisotropic emission of radiation in a CCSN, specifically through the emission of neutrinos. Jumping forward two decades, \cite{1996PhRvL..76..352B} and \cite{1997A&A...317..140M} proposed GW memory generation mechanisms through pulsar recoil and the anisotropic emission of energy in the form of neutrinos. The first three-dimensional CCSN simulation to include the GW emission sourced from neutrino emission was by \cite{2012A&A...537A..63M}. Moving quickly to the present, all models of CCSNe GW signals include (or have the ability to include) the GWs sourced from the neutrino fields. 

    From a detection stand point, the search for the CCSNe GW memory was often not included due to the presumed lack of sensitivity to frequencies associated with GW memory. However, recent work shows promise for detection of the memory -- more specifically the ramp up to the memory -- in current detectors -- E.G. see, \cite{2024PhRvL.133w1401R}. This is complemented with the development of new GW detectors with frequency sensitivities in the low-frequency range such as the millihertz detector Laser Interferometer Space Antenna (LISA; \cite{2019BAAS...51g..77T}) and the next generation terrestrial detectors Cosmic Explorer (CE; \cite{2019BAAS...51g..35R}) and the Einstein Telescope (ET; \cite{2020JCAP...03..050M}). Because of these advancements we seek to address low-frequency GWs through the window of CCSNe.

    This paper is organized as follows: Section \ref{sec:GW fron NU} walks though the derivation of \cite{1978ApJ...223.1037E} and \cite{1997A&A...317..140M}, shown in detail in \cite{2025PhRvD.112l3025R} and then combines the GWs sourced from the neutrino field and the fluid field. Section \ref{sec:detectors} discusses the detectability of CCSN GW memory in current detectors and provides predictions for GW memory investigations in future GW detectors including LISA, CE, and ET and reviews detection prospects for other proposed future detectors such as the proposed Lunar Gravitational-wave Antenna (LGWA; \cite{2025JCAP...01..108A}). 

\section{Gravitational Wave Emission from Core-Collapse Supernovae}
\label{sec:GW fron NU}
    Following the derivation of GWs sourced from anisotropic emission in \cite{2025PhRvD.112l3025R} we begin with the Einstein equations as presented in \cite{1973grav.book.....M},
    \begin{equation}
        G_{\mu \nu} = 8 \pi T_{\mu \nu}.
    \end{equation}
    We introduce a linear perturbation about Minkowski space, $h^{\mu \nu}$ and its ``trace-reversed'' form, $\bar{h}^{\mu \nu}$, and add to to the metric and rewrite the Einstein equations as a wave equation of the form,
    \begin{equation}
        \Box \bar{h}_{\mu \nu} = -16 \pi T_{\mu \nu}.
    \end{equation}
    This wave equation has a solution of the form,
    \begin{equation}
    \label{eq:pre_strain}
        \bar{h}_{\mu \nu}(x) = 4 \int d^{3} x' \frac{1}{|\vec{x} - \vec{x}'|}T_{\mu \nu}(x^{0}_{r}, \vec{x}').
    \end{equation}
    Following the formulation of \cite{1978ApJ...223.1037E} we write the stress-energy tensor in terms of a rate of energy loss, $\sigma(t')$, and the angular distribution of the emission, $f(t',\Omega')$,
    \begin{equation}
    \label{eq:stress}
        T_{i j}(x) = \frac{n_{i} n_{j}}{r^{2}} \int_{- \infty}^{\infty} dt' \sigma(t') f(t',\Omega') \delta(t - t' - r).
    \end{equation}
    We are able to rewrite the geometric factor in the denominator and arrive at the transverse-traceless GW strain for the general anisotropic emission of neutrino radiation energy in a system, 
    \begin{equation}
        \bar{h}_{i j}^{TT}(x) = 4 \int_{- \infty}^{t - r} \int_{4 \pi} d\Omega' dt' \frac{(n_{i} n_{j})^{TT}}{t - t' - r \cos{\theta}} \sigma(t') f(t',\Omega').
    \end{equation}
    It is from here that we follow \cite{1997A&A...317..140M} and expand the transverse-traceless geometry into a spherical polar observer frame and recover the two GW polarizations ($h_{+}$ and $h_{\times}$),
    \begin{equation}
    \label{eq:GWfromNu}
        \begin{split}
            h_{+}(x) & = 2 \int_{- \infty}^{t - r} \int_{4 \pi} d\Omega' dt' \frac{(1 - \cos^{2}{\theta}) \cos{2 \phi}}{t - t' - r \cos{\theta}} \sigma(t'), f(t',\Omega') \\
            h_{\times}(x) & = 2 \int_{- \infty}^{t - r} \int_{4 \pi} d\Omega' dt' \frac{(1 - \cos^{2}{\theta}) \sin{2 \phi}}{t - t' - r \cos{\theta}} \sigma(t') f(t',\Omega').
        \end{split}
    \end{equation}
    Turning to the directional and energy loss terms, $f(t',\Omega')$ and $\sigma(t')$, we approximate the directional loss of energy as the differential neutrino energy luminosity from a surface of 500 km from the core. This allows us to approximate,
    \begin{equation*}
        \frac{d L}{d \Omega'} (t', \Omega') \sim \sigma(t') f(t', \Omega').
    \end{equation*}
    As a specific example, in the \textsc{Chimera} models (\cite{2020ApJS..248...11B}), the differential luminosity, $\frac{d L}{d \Omega'}$, is defined in terms of the neutrino flux at a given radius,
    \begin{equation}
        \frac{d L_{E}^{\nu_{i}}}{d \Omega}(t,\Omega) = \frac{4 \pi r^{2} c}{(h c)^{3}} \frac{1}{\alpha^{4}} \int dE \psi^{(1)}(\phi, \theta, r, \nu_{i}, E) E^{3}.
    \end{equation}
    Here we also account for the general relativistic redshift effects through the lapse function $\alpha.$
    
\begin{figure}
        \centering
        \begin{subfigure}[t]{0.49\textwidth}
            \centering
            \includegraphics[width=\linewidth]{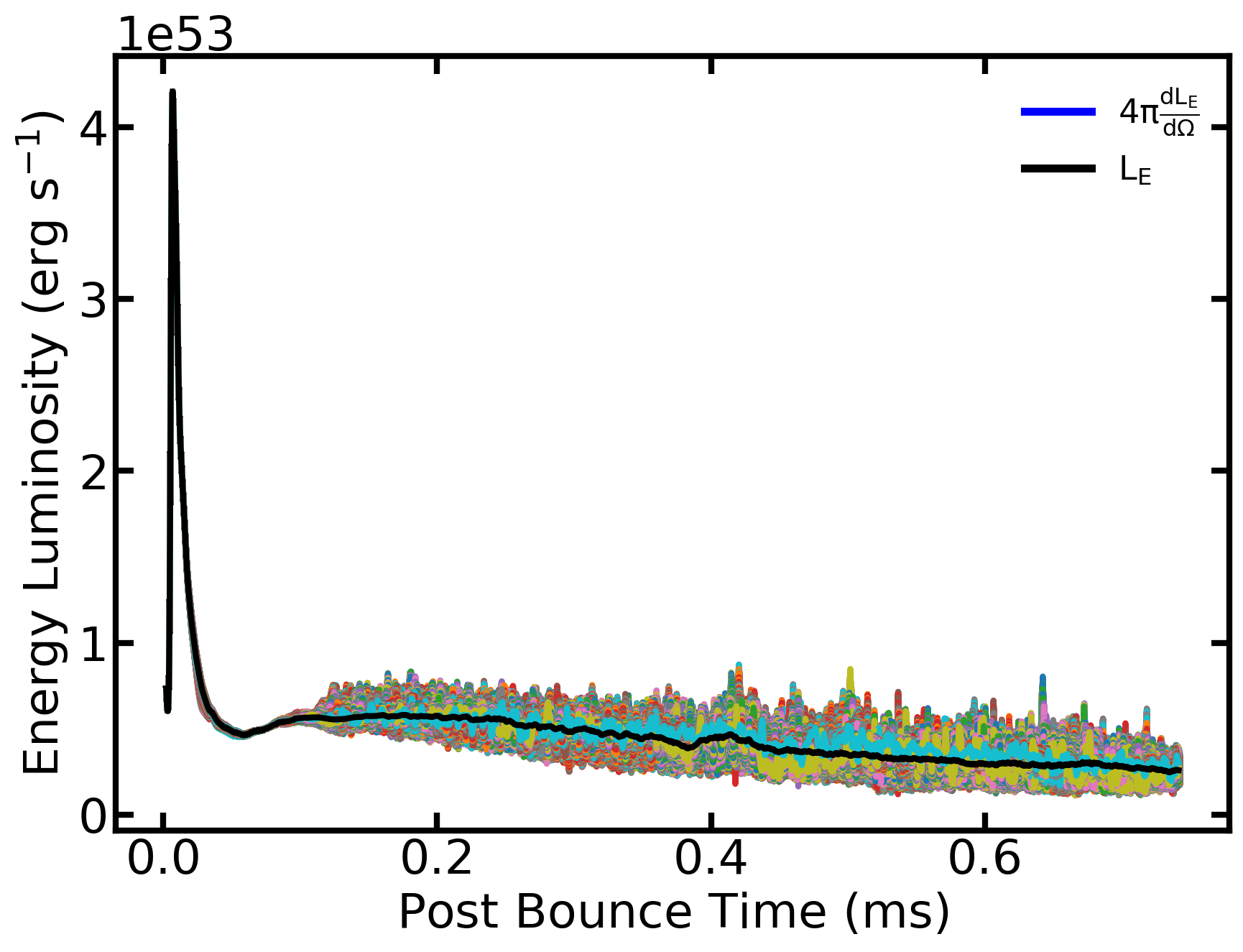} 
            \caption{Energy luminosity of the electron neutrinos.} \label{fig:diff_lum}
        \end{subfigure}
        \begin{subfigure}[t]{0.49\textwidth}
            \centering
            \includegraphics[width=\linewidth]{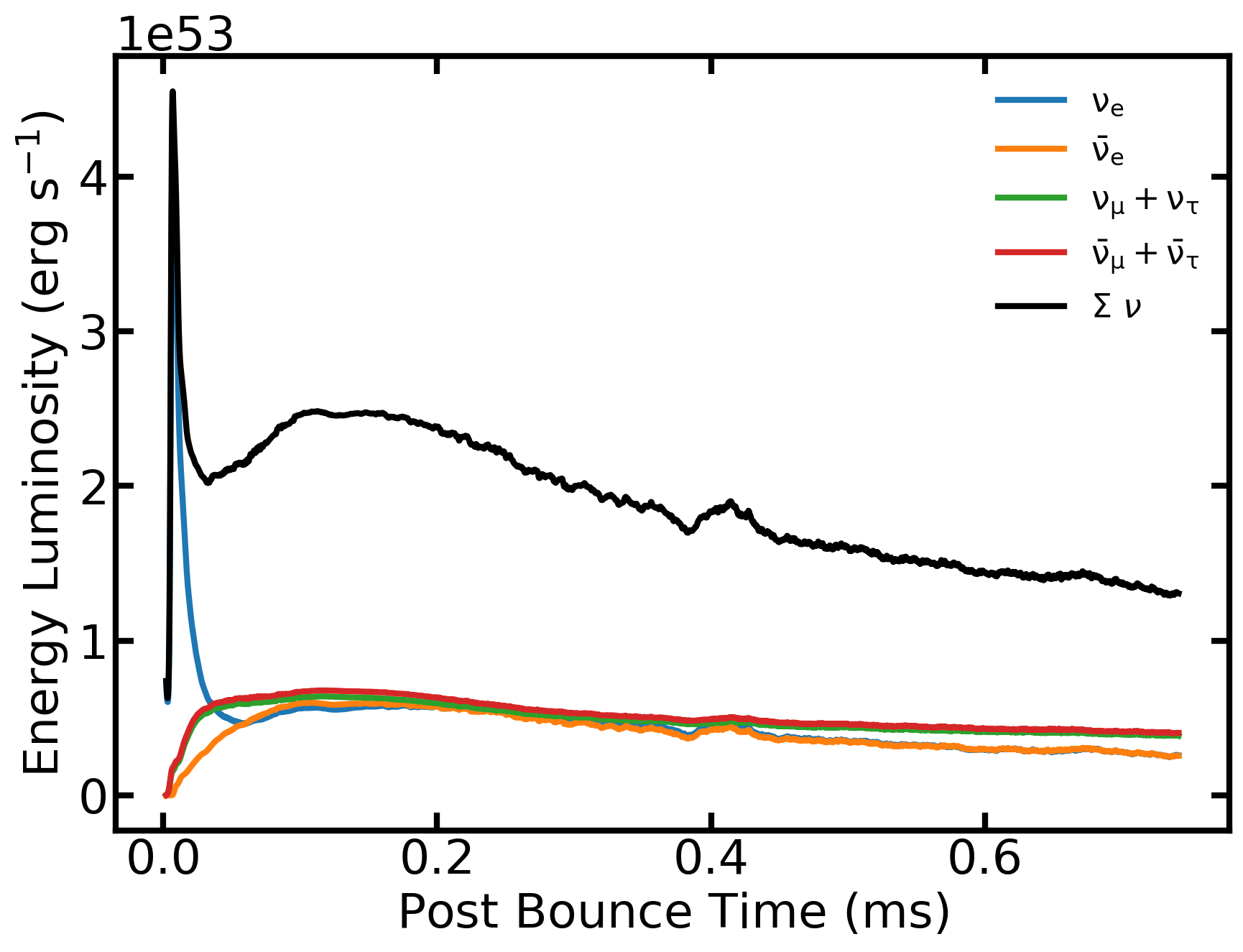} 
            \caption{Energy luminosity of all neutrino species and the total luminosity.} \label{fig:lum}
        \end{subfigure}
        \caption{Energy luminosity of neutrinos emitted at 500 km. In the left plot, the colored lines show $4 \pi$ times the differential luminosity and the black line shows the total luminosity emitted from a spherical surface at 500 km. In the right plot the colored lines show the luminosity from each species individually and the black line shows the total luminosity of neutrinos. Previously shown in \cite{2025PhRvD.112l3025R}.}
    \end{figure}

    Figs. \ref{fig:diff_lum} and \ref{fig:lum} show the neutrino energy luminosity for our fiducial D15-3D model, wherein Fig. \ref{fig:diff_lum} the variation in luminosity, after approximately 100 ms, is evident by the spread of the $4 \pi$ scaled differential luminosities (in color) around the $\mathrm{4 \pi}$-integrated luminosity (in black). While all rays follow the same general trend due to accretion events onto the surface of the proto-neutron star and turbulent effects at the 500 km radius of extraction there are noticeable peaks on that are up to twice as large as the integrated quantity. These variations lead to the anisotropy of emission and the source of the GW emission from the neutrino field. Fig. \ref{fig:lum} shows the integrated energy luminosity of all four neutrino species evolved in \textsc{Chimera}, note that $\nu_{x} = \nu_{\mu} + \nu_{\tau}$ and $\bar{\nu}_{x} = \bar{\nu}_{\mu} + \bar{\nu}_{\tau}$, and the total luminosity of all combined neutrino fields. While weak processes and neutrino detection rely on the individual neutrino fields, for the GW sourcing we need the entire radiation field's stress-energy tensor and as such the total neutrino luminosity and its differential variations. Here the overall shape is dominated by the electron neutrinos, but the amplitude and features of the luminosity are enhanced by the inclusion of the other fields.
    Using the integrated luminosity passing through our 500 km sphere, $L^{\nu}_{E}(t) = \int_{4 \pi} d \Omega' \frac{d L^{\nu}} {d\Omega'}(\Omega', t')$, and angular weights for the geometric factors coupled to the differential luminosity, we write the gravitational wave strain as
    \begin{equation}
        h^{\nu_{i}}_{+/\times}(t, \alpha,\beta) = \frac{2 G}{c^{4} r} \int_{-\infty}^{t - r/c} dt'  \frac{1}{L^{\nu_{i}}_{E}(t)} \int_{4 \pi} d \Omega' W_{+/\times}(\alpha, \beta, \Omega') \frac{d L^{\nu_{i}}_{E}} {d\Omega'}(\Omega', t').
    \end{equation}
    We also introduce an anisotropy parameter to detail how anisotropic the neutrino emission is, given a particular viewing direction, 
    \begin{equation}
    \label{eq:Anisotropy}
        \alpha^{\nu_{i}}_{+/\times}(t, \alpha, \beta) = \frac{1}{L^{\nu_{i}}_{E}(t)} \int_{4 \pi} d \Omega' W_{+/\times}(\alpha, \beta, \Omega') \frac{d L^{\nu_{i}}_{E}} {d\Omega'}(\Omega', t').
    \end{equation}

\begin{figure}
        \centering
        \begin{subfigure}[t]{0.49\textwidth}
            \centering
            \includegraphics[width=\linewidth]{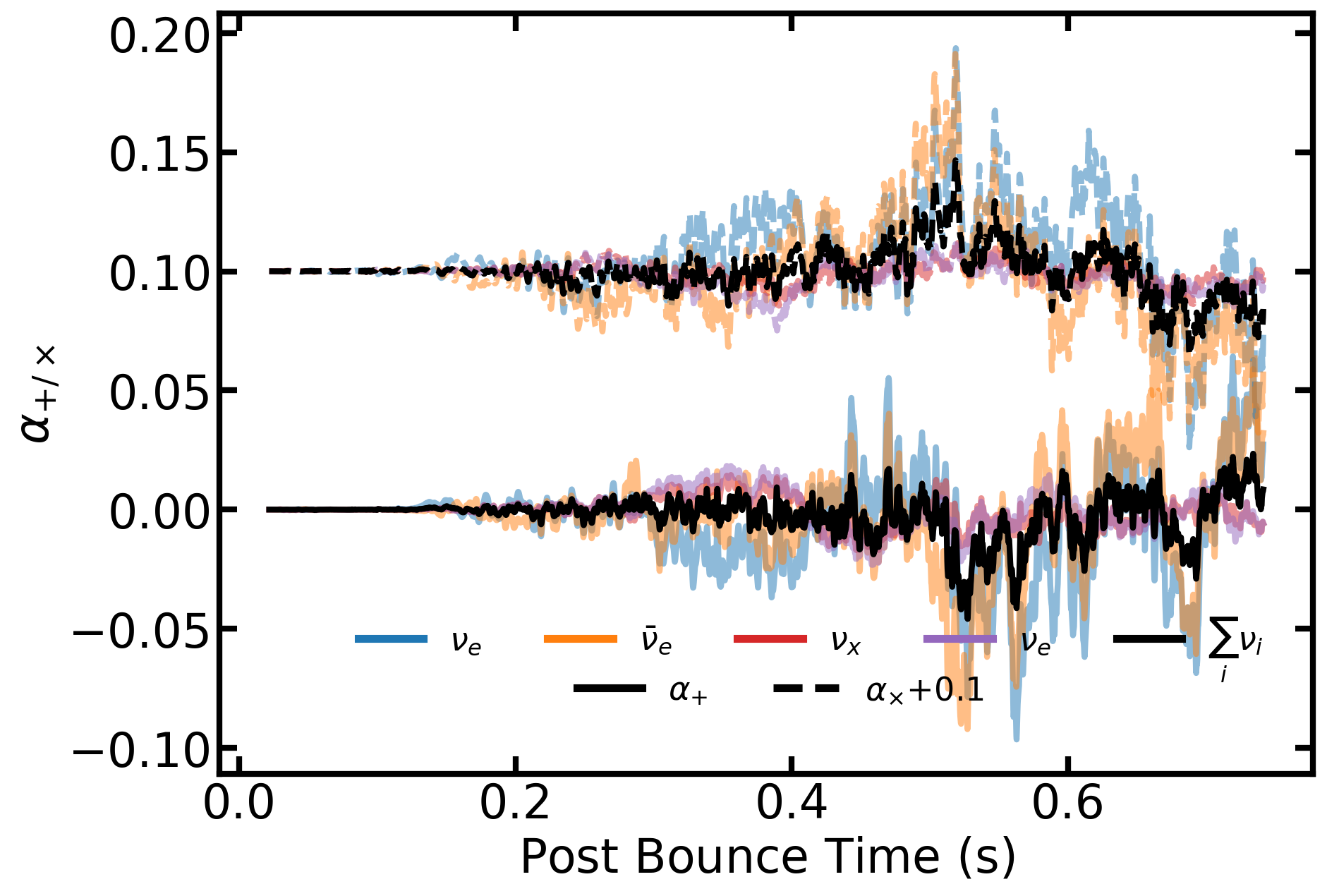} 
            \caption{Plus and cross polarization neutrino luminosity anisotropy.} \label{fig:anisotropy}
        \end{subfigure}
        \begin{subfigure}[t]{0.48\textwidth}
            \centering
            \includegraphics[width=\linewidth]{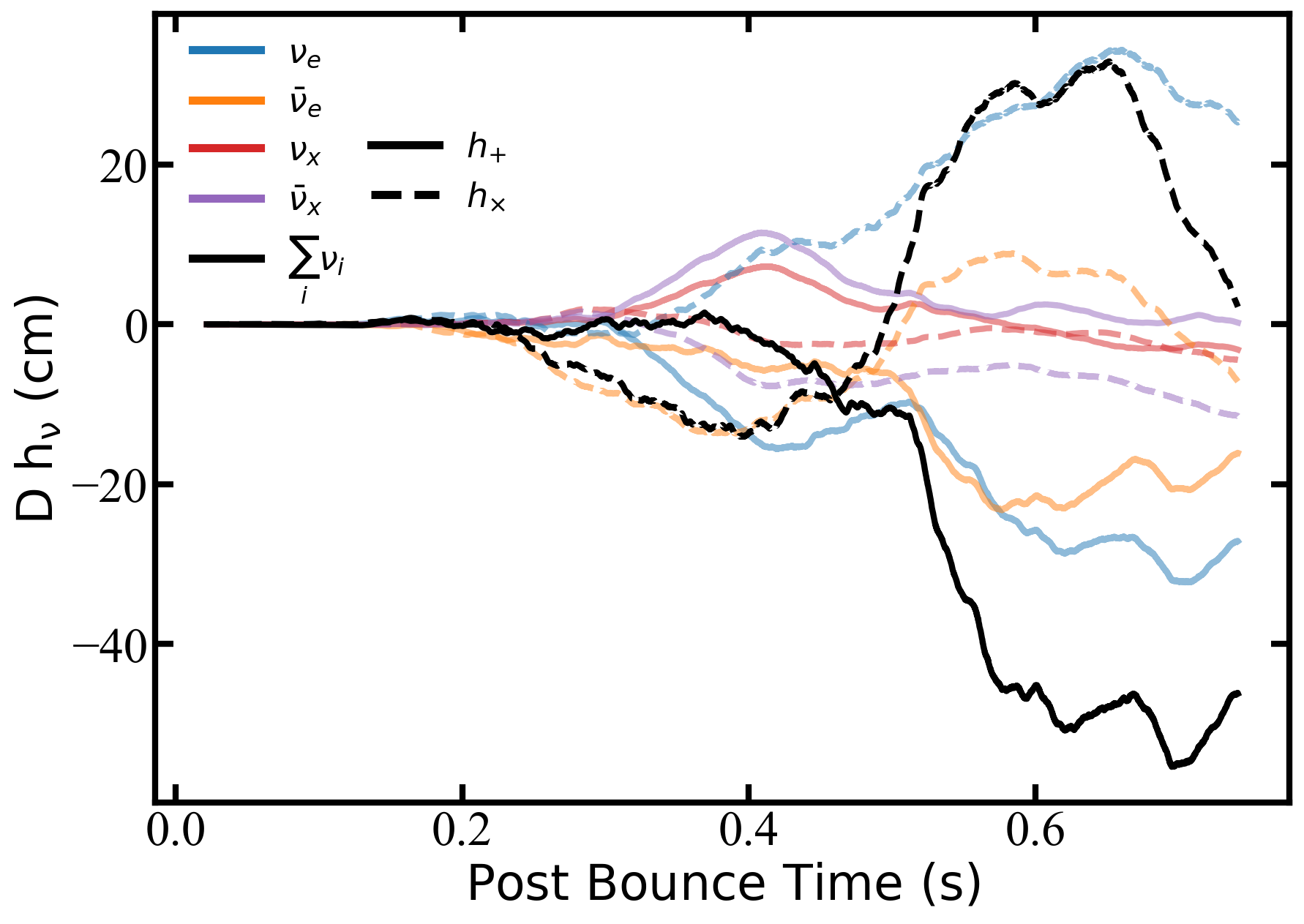} 
            \caption{Plus and cross polarization gravitational wave strain sourced from neutrinos.} \label{fig:nu strain}
        \end{subfigure}
        \caption{For both the left and right plots, the colored lines indicate the anisotropy (left) and GW strain (right) from the individual neutrino species. The black lines indicate the total anisotropy and strain, respectively. The solid lines indicate the plus polarization and the dashed lines indicate the cross polarization. Previously shown in \cite{2025PhRvD.112l3025R}, but highlighting a different viewing orientation with respect to the source.}
    \end{figure}

    With this new anisotropy parameter we can then condense the gravitational wave strain as,
    \begin{equation}
        h^{\nu_{i}}_{+/\times}(t, \alpha,\beta) = \frac{2 G}{c^{4} r} \int_{-\infty}^{t - r/c} dt' L^{\nu_{i}}_{E}(t) \alpha^{\nu_{i}}_{+/\times}(t, \alpha, \beta).
    \end{equation}
    The anisotropy parameter and the GW strain sourced from the neutrino luminosity is shown in Figs. \ref{fig:anisotropy} and \ref{fig:nu strain} respectively. As can bee seen by comparing Figs. \ref{fig:anisotropy} and \ref{fig:nu strain}, a positive anisotropy will lead to a positive GW strain and a negative anisotropy will lead to a negative GW strain. The plus-polarization (solid black line) exhibits an over all negative trend; The anisotropy remains negative until approximately 400 ms, after which it returns to zero and the GW strain remains constant. This is followed by a larger negative anisotropy and a negative strain, before both turn over at around 650 ms. Given the form of the GW strain sourced from the neutrino anisotropy, it is apparent that even if the anisotropy goes to zero, the GW strain will not, indicating that the final value of the GW memory will be dominated by the neutrino asymmetries.

    With the GWs sourced from the neutrino field defined, we can also investigate the GWs sourced from the fluid field. These have been investigated in detail in many of the studies previously mentioned and specifically for the model shown here, the extraction of these waveforms is detailed in \cite{2023PhRvD.107d3008M}. The final expression comes from the quadrupole moment of the transverse-traceless gravitational wave strain (\cite{2006RPPh...69..971K}), 
    \begin{equation}
            h_{+} = \frac{h_{\theta \theta}^{TT}}{r^{2}},
    \end{equation}
    \begin{equation}
            h_{\times} = \frac{h_{\theta \phi}^{TT}}{r^{2} \sin{\theta}},
    \end{equation}
    where
    \begin{equation}
            h_{i j}^{TT} = \frac{G}{c^{4}} \frac{1}{r} \sum_{m = - 2}^{+2} \frac{d^{2} I_{2 m}}{d t^{2}} \bigg( t - \frac{r}{c} \bigg) f_{i j}^{2m}.
    \end{equation}
    Where the second time derivative of the mass quadrupole moment, $I$ is expanded using spherical harmonics, where $f^{2m}_{ij}$ are the tensor spherical harmonics and $i$ and $j$ run over $r$, $\theta$, and $\phi$
    \begin{equation}
        f_{i j}^{l m} = r^{2} 
        \begin{pmatrix}
            0 & 0 & 0 \\
            0 & W_{l m} & X_{l m} \\
            0 & X_{l m} & -\sin^{2}{\theta} W_{l m}
        \end{pmatrix}, 
    \end{equation}
    where
    \begin{equation}
        X_{lm} = 2 \frac{\partial}{\partial \phi} \left( \frac{\partial}{\partial \theta} - \cot{\theta} \right) Y_{l m}(\theta, \phi)
    \end{equation}
    and
    \begin{equation}
        W_{l m} = \left( \frac{\partial^{2}}{\partial \theta^{2}} - \cot{\theta} \frac{\partial}{\partial \theta} - \frac{1}{\sin^{2}{\theta}} \frac{\partial^{2}}{\partial \phi^{2}} \right) Y_{l m}(\theta, \phi)
    \end{equation}
    and $Y_{l m}$ are the spherical harmonics.
    
\begin{figure}
        \centering
        \begin{subfigure}[t]{0.48\textwidth}
            \centering
            \includegraphics[width=\linewidth]{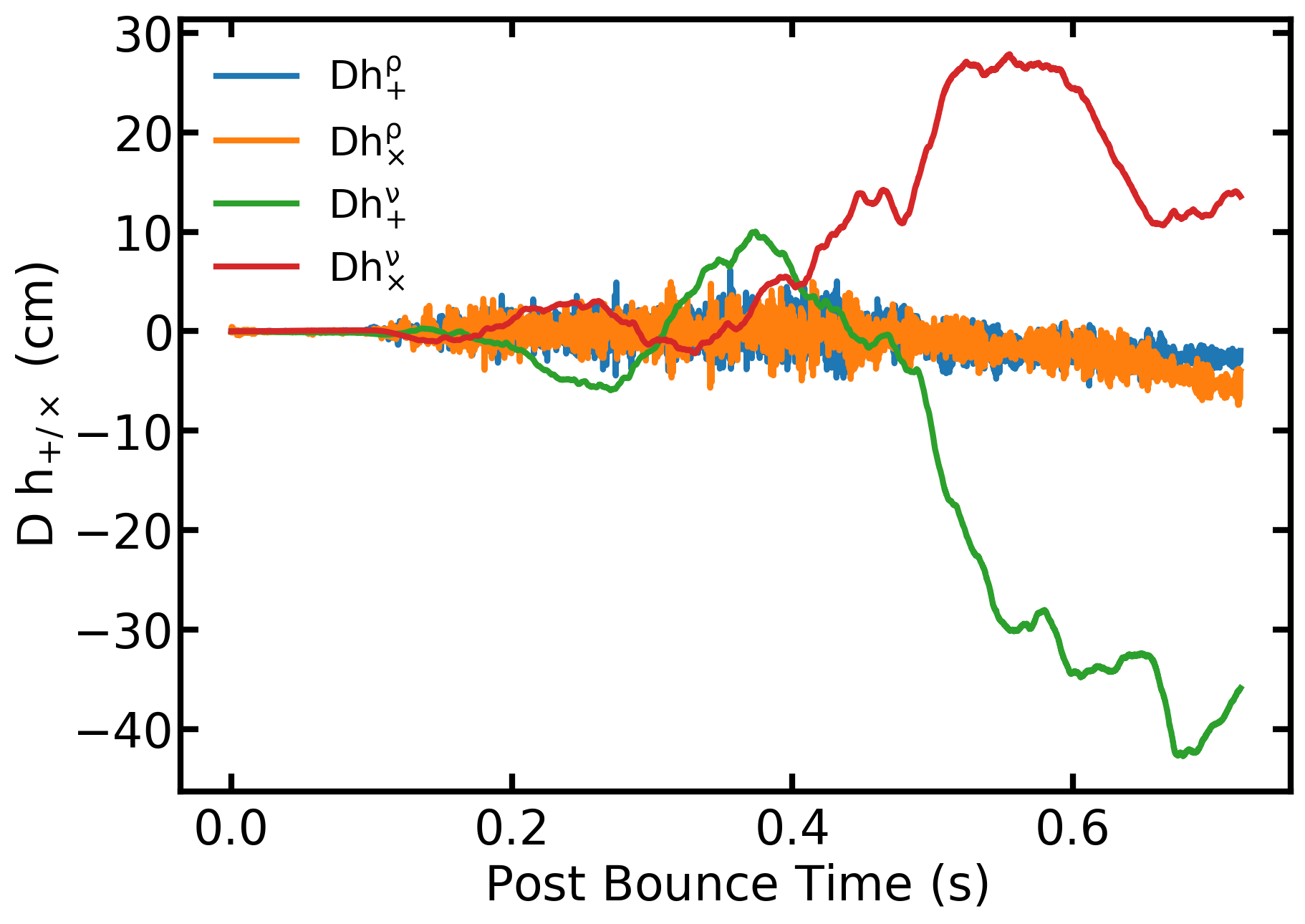} 
            \caption{Gravitational wave strain sourced from both the neutrino and fluid fields.} \label{fig:strain}
        \end{subfigure}
        \begin{subfigure}[t]{0.49\textwidth}
            \centering
            \includegraphics[width=\linewidth]{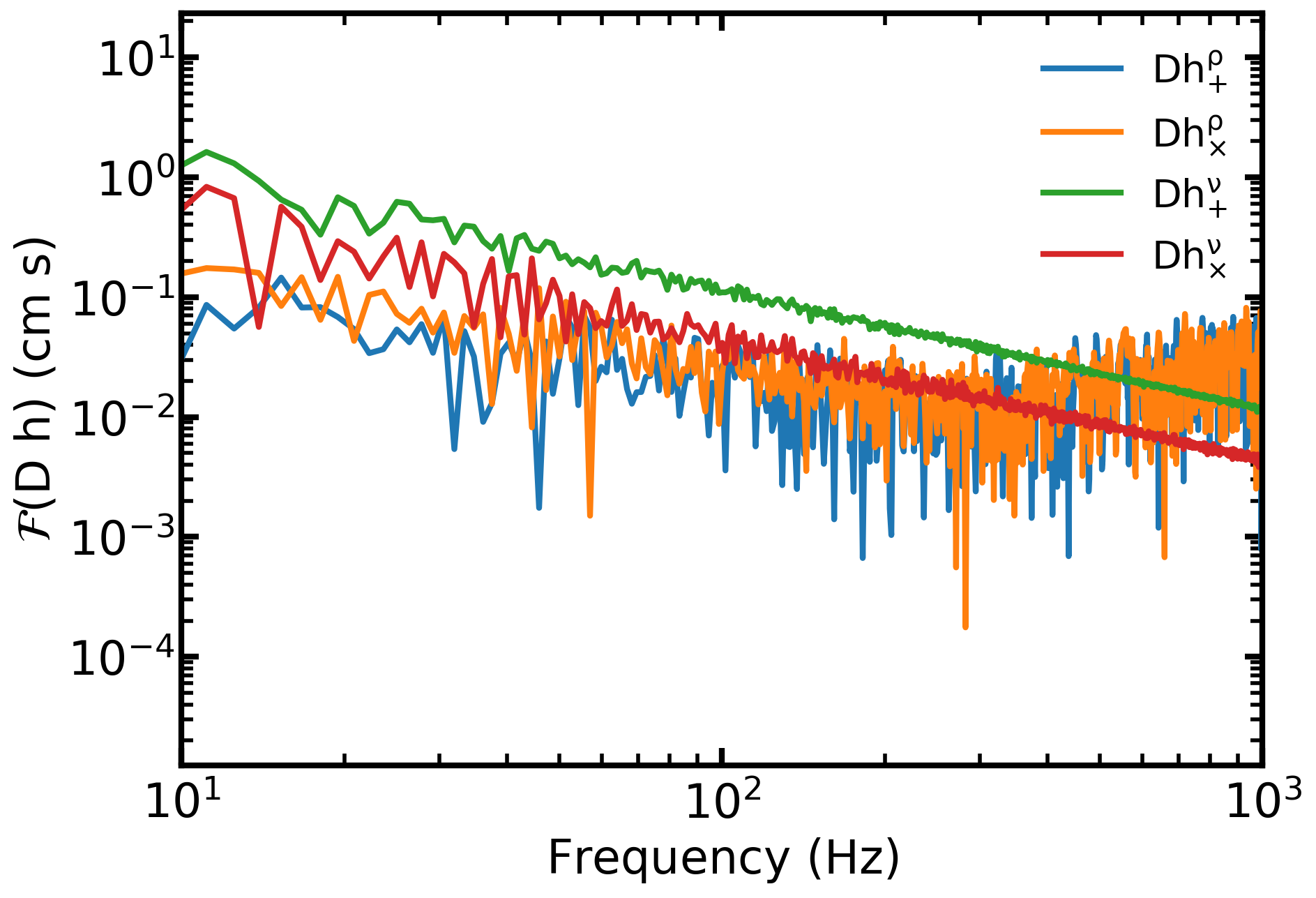} 
            \caption{Fourier transform of the gravitational wave strain sourced from both the neutrino and fluid fields.} \label{fig:fft}
        \end{subfigure}
        \caption{In both the strain, $\mathrm{D \ h}$, (left) and Fourier transform of the strain, $\mathrm{\mathcal{F}(D \ h)}$, (right) the plus and cross polarizations for both the neutrino sourced and fluid sourced GWs are shown. Both are scaled to the source.}
    \end{figure}

    The GWs sourced from the fluid and the neutrino fields are shown in Fig. \ref{fig:strain}, where it is immediately apparent that the overall amplitude of the fluid-sourced GWs ($\mathrm{h_{+/\times}^{\rho}}$) is smaller than that of the neutrino-sourced GWs, but the fluid-sourced GWs have a much higher frequency indicating that the energy released in the form of gravitational radiation is dominated by the matter fluid components. This article is focused on the low-frequency component of the GWs, which although less energetic have a higher amplitude. Fig. \ref{fig:fft} shows the Fourier transform of both $\mathrm{h_{+/\times}^{\rho}}$ and $\mathrm{h_{+/\times}^{\nu}}$ and in the frequency range of interest, $\mathrm{f < 50}$ Hz, the neutrino-sourced GWs are dominant over the fluid-sourced GWs. 

\begin{figure}
        \centering
        \begin{subfigure}[t]{0.49\textwidth}
            \centering
            \includegraphics[width=\linewidth]{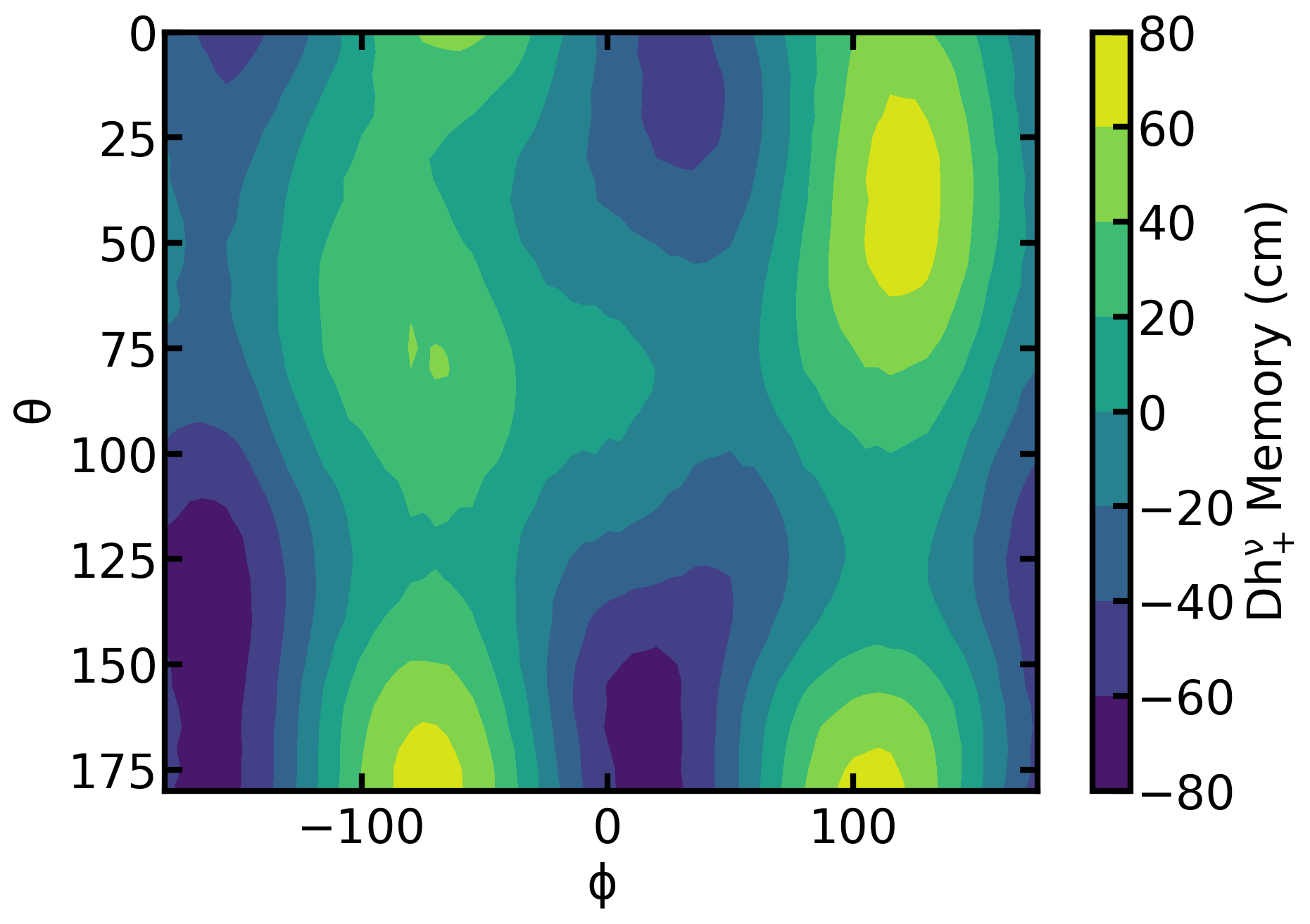} 
            \caption{Plus polarization GWs sourced from the neutrino field.} \label{fig:hnp_m}
        \end{subfigure}
        \begin{subfigure}[t]{0.49\textwidth}
            \centering
            \includegraphics[width=\linewidth]{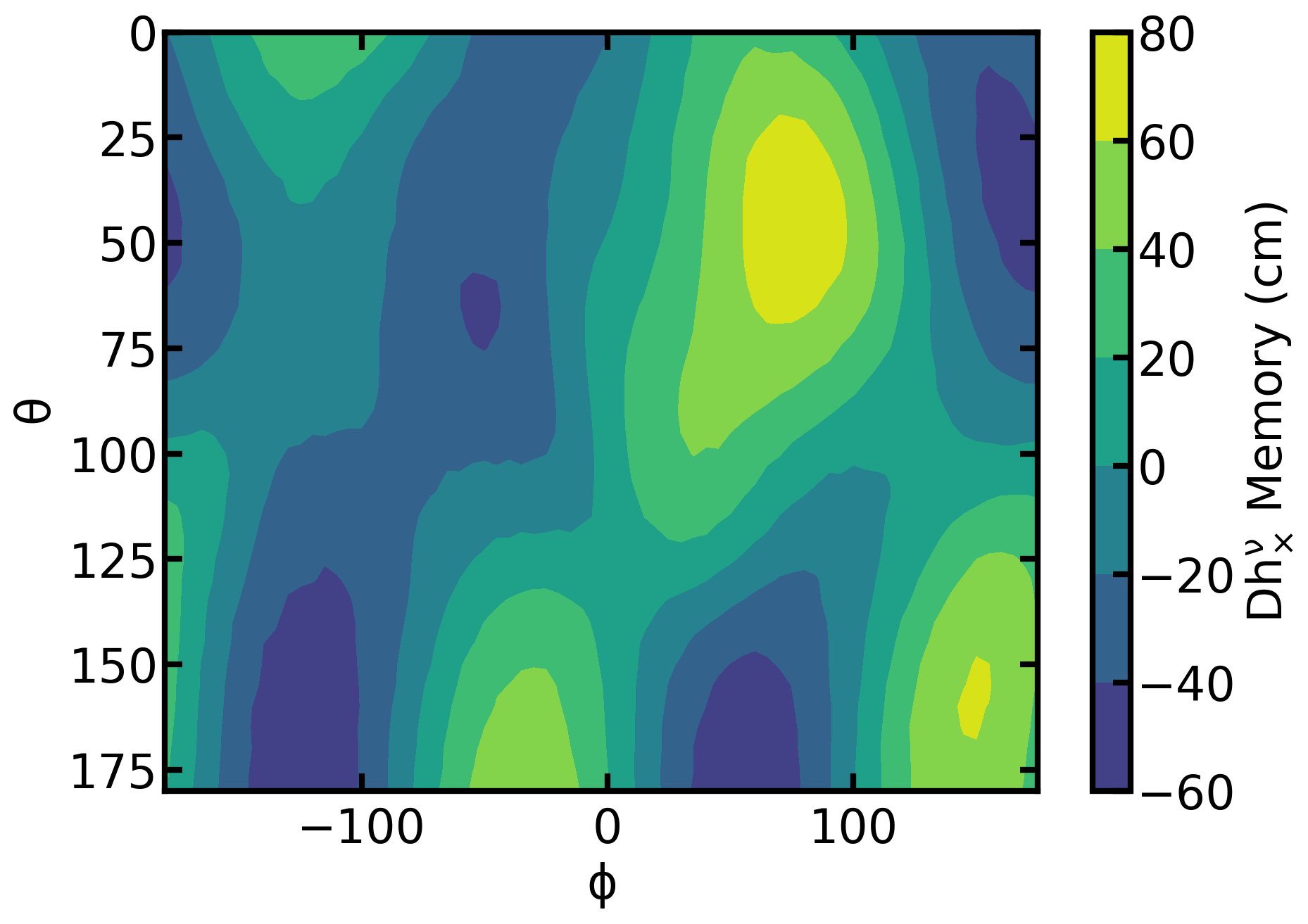} 
            \caption{Cross polarization GWs sourced from the neutrino field.} \label{fig:hnc_m}
        \end{subfigure}
        \begin{subfigure}[t]{0.49\textwidth}
            \centering
            \includegraphics[width=\linewidth]{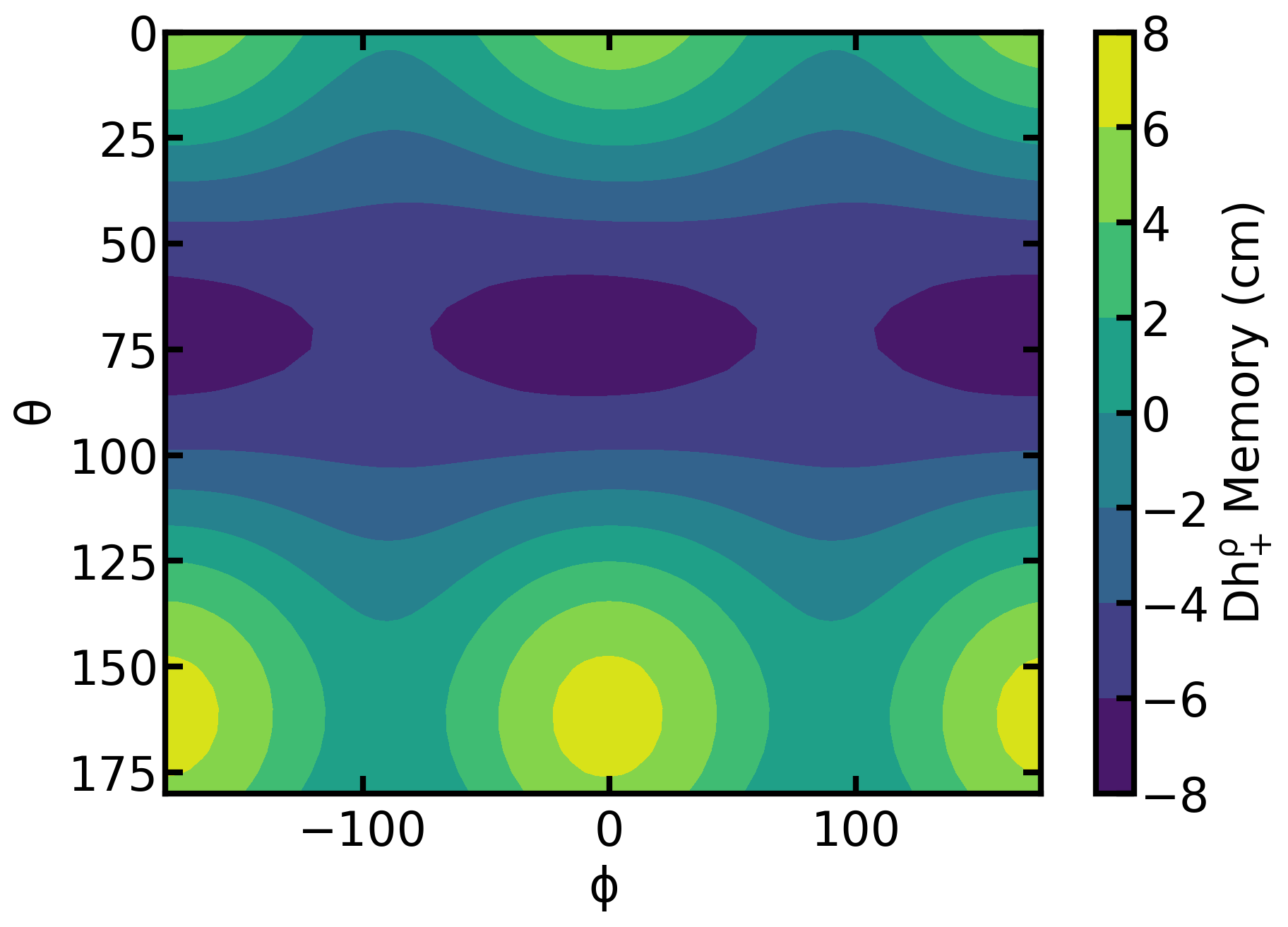} 
            \caption{Plus polarization GWs sourced from the fluid field.} \label{fig:hfp_m}
        \end{subfigure}
        \begin{subfigure}[t]{0.49\textwidth}
            \centering
            \includegraphics[width=\linewidth]{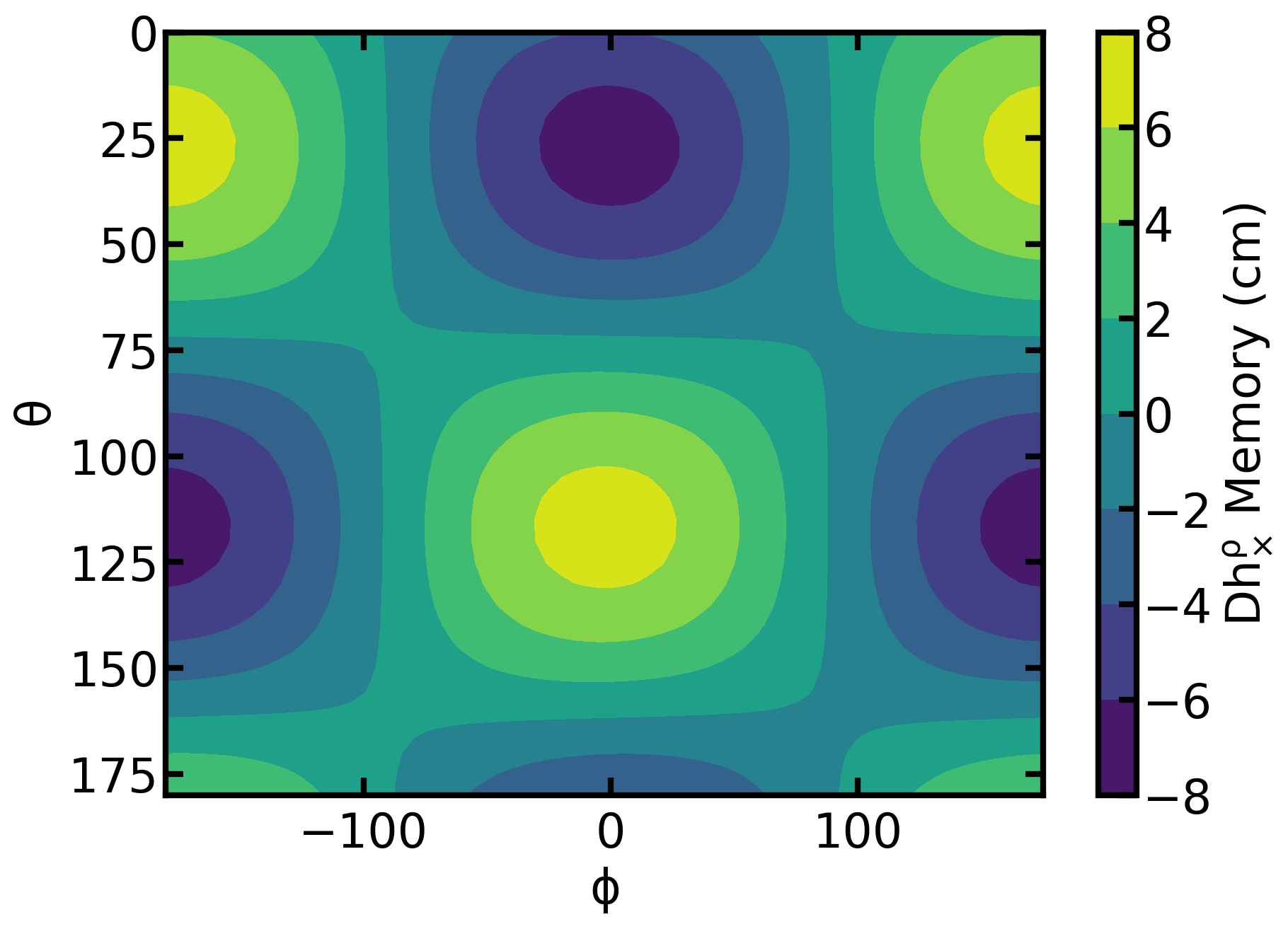} 
            \caption{Cross polarization GWs sourced from the fluid field.} \label{fig:hfc_m}
        \end{subfigure}
        \caption{Gravitational wave strain from the last time step of the model. In all sub-figures the y-axis is the polar angle and the x-axis is the azimuthal angle coincident with the spherical-polar grid of the computational domain.}
    \end{figure}

\begin{figure}
        \centering
        \begin{subfigure}[t]{0.48\textwidth}
            \centering
            \includegraphics[width=\linewidth]{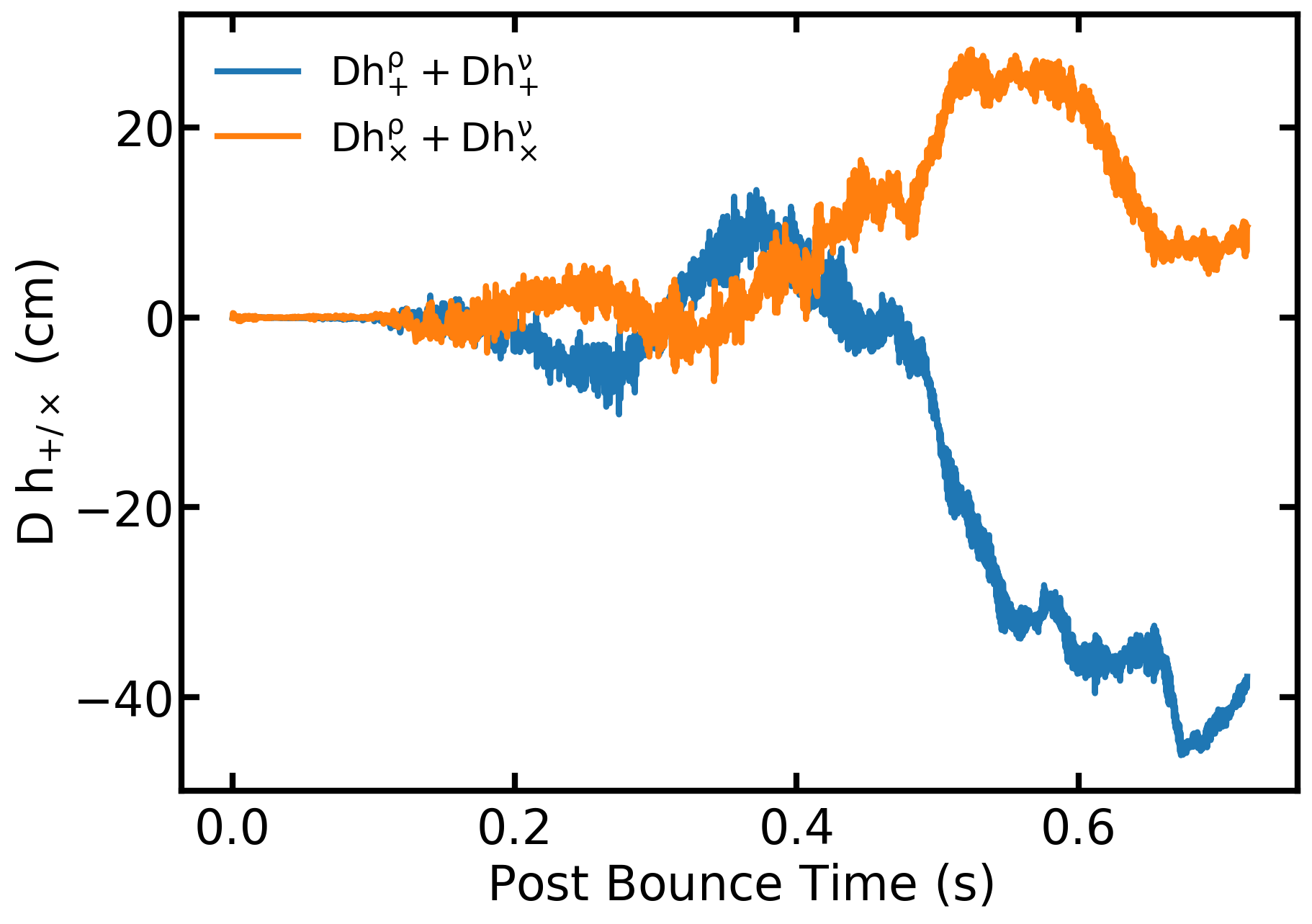} 
            \caption{Gravitational wave strain.} \label{fig:total strain}
        \end{subfigure}
        \begin{subfigure}[t]{0.49\textwidth}
            \centering
            \includegraphics[width=\linewidth]{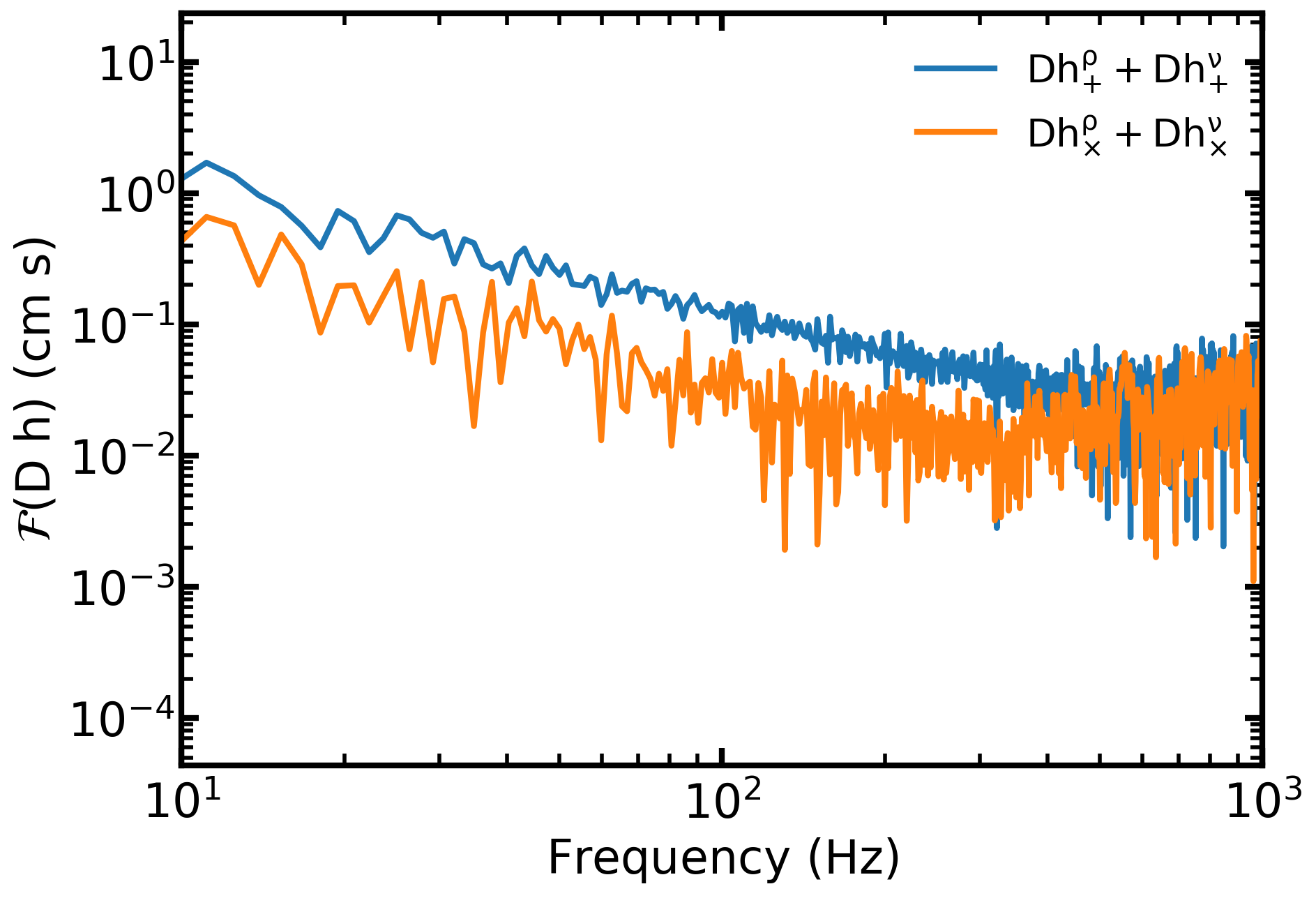} 
            \caption{Fourier transform of the gravitational wave strain.} \label{fig:total fft}
        \end{subfigure}
        \caption{In both the strain, $\mathrm{Dh}$, (left) and the Fourier transform of the strain, $\mathrm{\mathcal{F}(Dh)}$ (right) the plus (purple) and cross (magenta) polarizations are shown. Both figures plot the GW signal scaled to the source.}
    \end{figure}

    If we look at the final amplitude of the individual strains, see Figs. \ref{fig:hnp_m}, \ref{fig:hnc_m}, \ref{fig:hfp_m}, and \ref{fig:hfc_m}, we can see the global morphology of both the fluid and neutrino fields as viewed away from the source. In the case of our D15-3D model, asymmetry in the explosion (GWs sourced from the geometry of the expanding matter) is different from that of the neutrino field. There are locations where the memory associated with the matter and with the neutrinos will add constructively and locations where the will add destructively. We note that our signals stop before the GW emission is no longer evolving and long-time simulations such as those shown in \cite{2024ApJ...975...12C} suggest that the signal will still evolve over second timescales. From models that have ``completed'', such as the D9.6-3D model shown in \cite{2023PhRvD.107d3008M}, the signal will saturate to a constant value. Given both sources we can linearly combine them to investigate the total signal as shown in Figs. \ref{fig:total strain} and \ref{fig:total fft}. These figures show that the overall GW morphology is dominated by the neutrino-sourced signal, but there is a higher-frequency, smaller-amplitude ``embedded'' signal from the matter.
    
\begin{figure}
        \centering
        \includegraphics[width=0.9\linewidth]{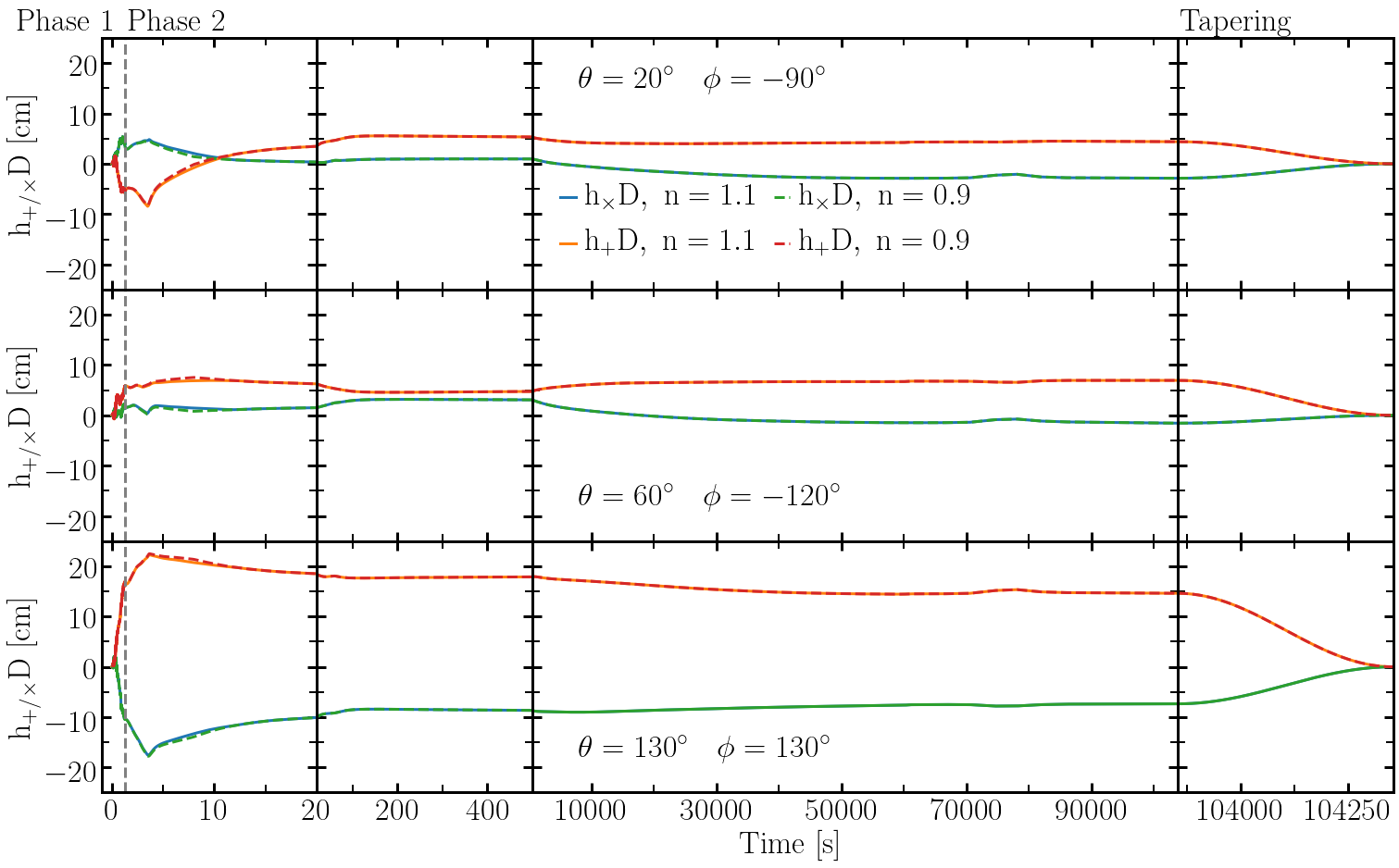}
        \caption{The different vertical panels show GW signals from different observer orientations. The blue and green dashed lines plots the cross polarization strains assuming an exponential factor of 1.1 and 0.9 respectively. The orange and red dashed lines plots the plus polarization strains assuming an exponential factor of 1.1 and 0.9 respectively. This figure is reproduced from \cite{2022PhRvD.105j3008R}.}
        \label{fig:nu gw extension}
    \end{figure}

    Due to the cost of running extended models (longer than approximately 1 supernova second), if CCSN models are extended to shock-breakout, or beyond, many approximations have to be made, chief among them is the removal of neutrino transport in favor of more simplified energy loss and PNS cooling methods (see \cite{2012A&A...537A..63M}, \cite{2021ApJ...921..113S}, and \cite{2025arXiv250916314V} for examples). As such, we cannot source GW signals from the neutrino fields. \cite{2022PhRvD.105j3008R} propose a method to extend the GW signal sourced from the neutrino field. From spherically symmetric simulations (see for example \cite{1999ApJ...513..780P, 2001ApJ...562..887T}), we know that the total neutrino luminosity during the cooling phase of the PNS is well approximated by an exponential function,
    \begin{equation}
        \label{eq:lum1.3}
        L_{\nu}^{E}(t) = C t^{-n},
    \end{equation}
    with $n \sim 1$. Assuming that the accretion rate has dropped significantly, the neutrino luminosity should be well approximated by Eq. \ref{eq:lum1.3}. Given this form of the luminosity we can determine the constant $C$ by requiring that the luminosity from the robust neutrino transport at the end of the model ($t_{end}$) be continuous with this extension; i.e., that
    \begin{equation}
        C = t_{end}^{n} L_{\nu}(t_{end}).
    \end{equation}
    We can now take a pragmatic approach to approximating this signal by assuming that the anisotropy parameters remain constant at the end of the simulation, and we can write the extended GW signal sourced from the neutrino anisotropy as
    \begin{equation}
    \label{eq:gw_ex1}
        \begin{split}
        h_{\times/+}^{ext}(t_{ext}) &= h_{\times/+}^{end} + \frac{2G}{c^4 R} \alpha_{\times/+}^c\int_{t_{end}}^{t_{ext}} C \tau^{-n} (\tau) \mathrm{d}\tau \nonumber \\ 
        & = h_{\times/+}^{end} + \frac{2G}{c^4 R} \frac{\alpha_{\times/+}^c C}{1-n} \bigg(t_{ext}^{1-n}-t_{end}^{1-n} \bigg).
        \end{split}
    \end{equation}
    Here $h_{\times/+}^{end}$ represents the GW signal at the time step before the neutrino transport is turned off, $\alpha_{\times/+}^c$ represents the constant anisotropy parameters, and $t_{ext}$ is the extended time. Fig. \ref{fig:nu gw extension} is reproduced from \cite{2022PhRvD.105j3008R} and shows an example of the extended GW signal sourced from the neutrino field assuming a constant anisotropy from the W15-2 model of the \cite{2012A&A...537A..63M} simulation suite. Phase 1 represents the portion of the signal when neutrino transport is included, and Phase 2 represents the portion of the signal that is extended. The signals that are showcased are the total (matter and neutrino sourced) strains for three different viewing angles and two different exponential factors, $n$. We can readily see that varying $n$ around 1 does not alter the signal much and that in all cases the signal reaches a maximum or minimum well before shock breakout and then remains roughly constant. This extension is further motivated by long time models that include neutrino transport, like \cite{2024ApJ...975...12C}, which show the evolution of the GWs sourced from the neutrino anisotropy evolving on the order of a few seconds before approaching some non-zero amplitude.  

    This deterministic evolution from a zero to a non-zero signal motivates the use of a template in searching for these low-frequency signals in current and future data. The emission associated with the linear-memory ramp-up, emission below 100 Hz, can be fit with a logistic function proposed by \cite{2022PhRvD.105j3008R}, given by
    \begin{equation}
    \label{eq:logistic_fit}
        f(t,t_{0},k,L) = \frac{L}{1+e^{-k (t - t_{0})}},
    \end{equation}
    where $t_{0}$ is the center of the rise time, $k$ is the frequency of the memory ramp-up, and $L$ is the memory amplitude. In addition to $k$, we define $\tau = \frac{1}{k}$, which describes the timescale associated with the rise to the memory. Note that we can always shift our signal and our associated fit such that the central time is defined to be zero, $\mathrm{t' = t-t_{0}}$. We can fit this template to our signal across all observer orientations and define a characteristic ramp-up frequency, which for the data presented here is $k_{+} = 56.18 \pm 51.46$ and $k_{\times} = 61.06 \pm 52.16$ Hz, discussed in more detail in \cite{2025PhRvD.112l3025R}.

\section{Gravitational Wave Memory in Current and Future Detectors}
\label{sec:detectors}

\begin{figure}
        \centering
        \begin{subfigure}[t]{0.47\textwidth}
            \centering
            \includegraphics[width=\linewidth]{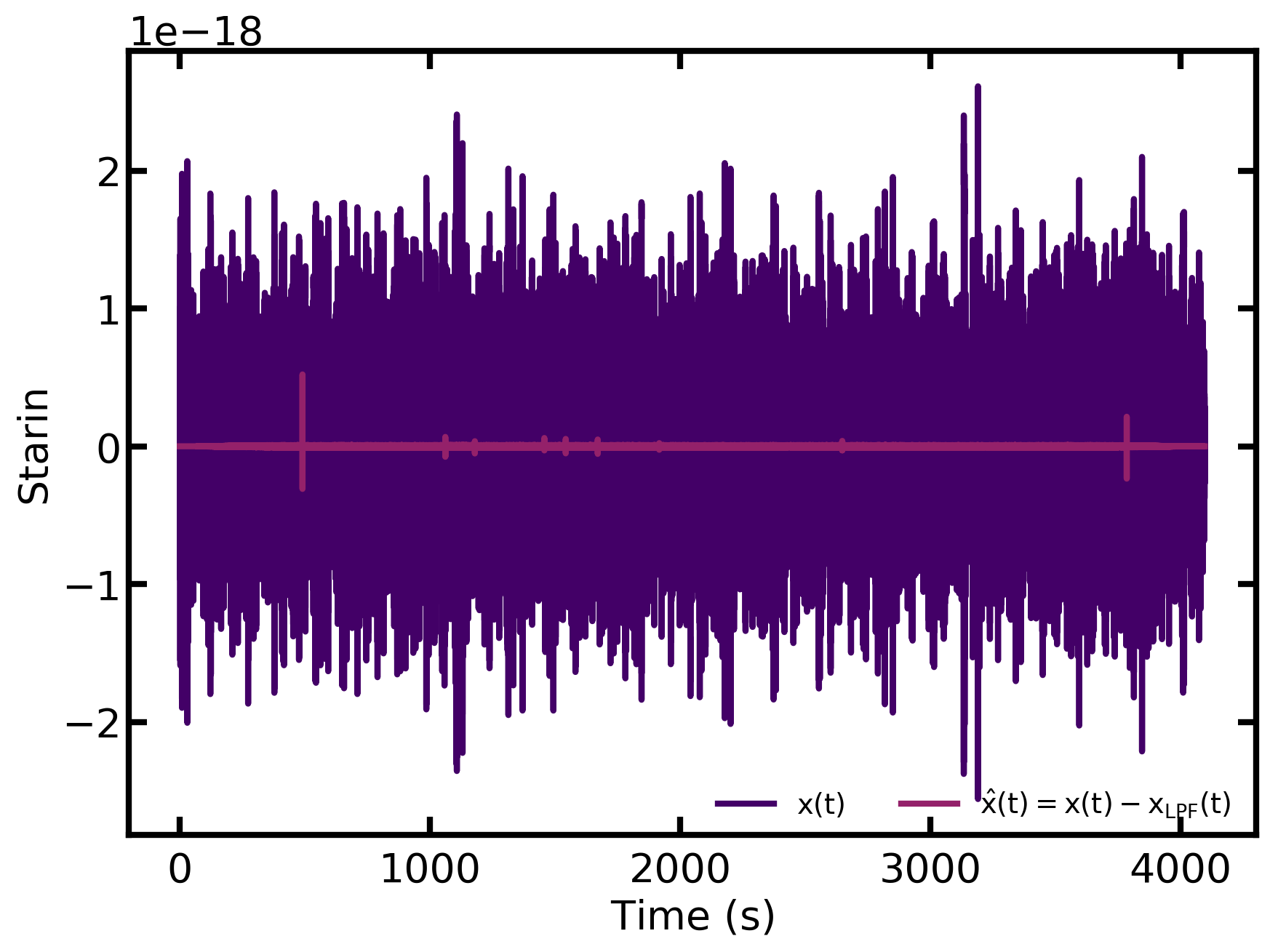} 
            \caption{Sample of real LIGO noise from the O3b run at the Livingston site.} \label{fig:L1}
        \end{subfigure}
        \begin{subfigure}[t]{0.5\textwidth}
            \centering
            \includegraphics[width=\linewidth]{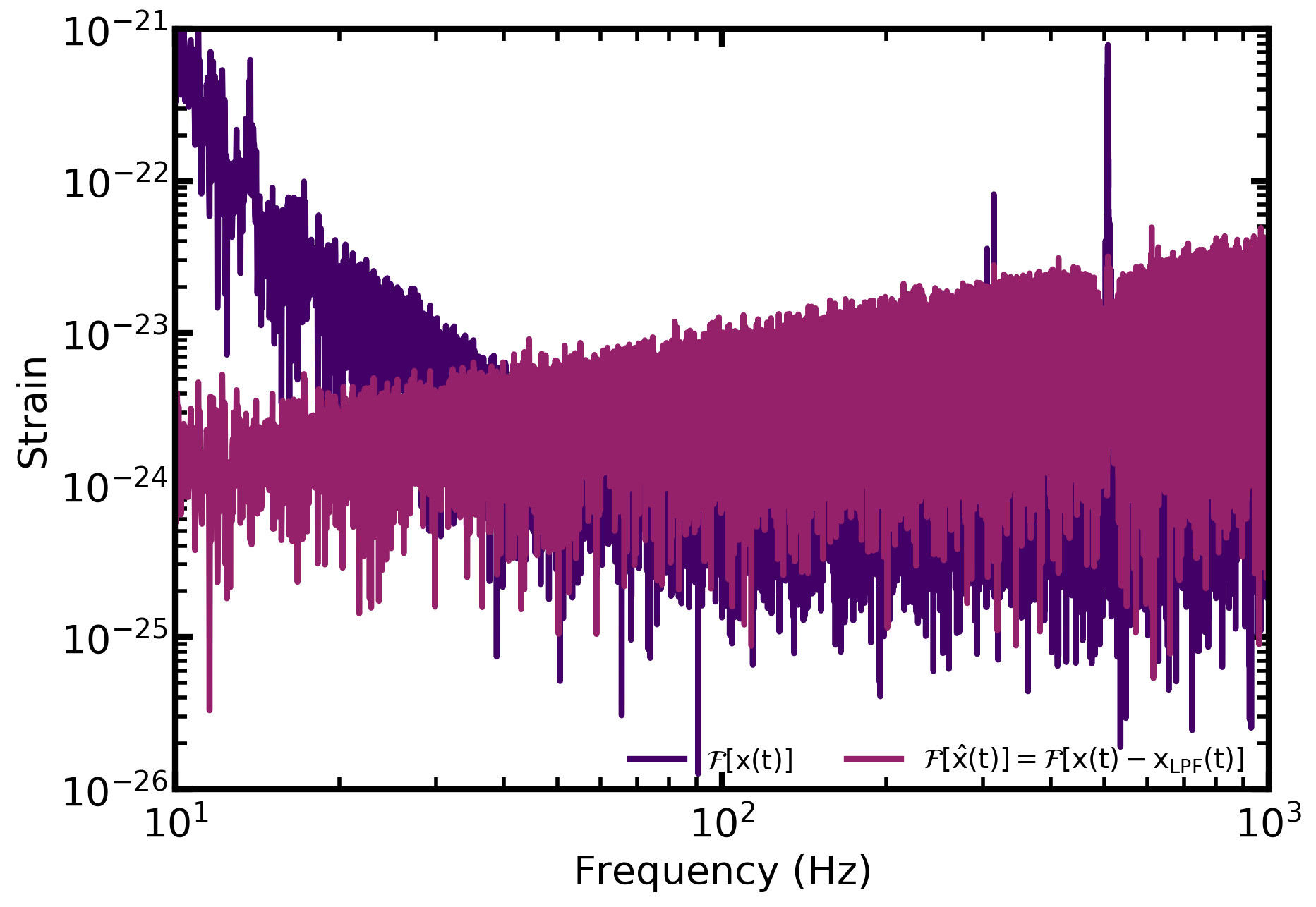} 
            \caption{Amplitude spectral density of real LIGO noise from the O3b run at the Livingston site.} \label{fig:L1_ASD}
        \end{subfigure}
        \caption{The linear predictive filter, when applied to this sample of Livingston O3b data with GPS time 1262178304, reduces the overall amplitude of the time series data as shown in the left plot. The right plot shows the same time series in frequency space.}\label{fig:real livingston}
    \end{figure}

    Given this template and the total GW signal arriving at a detector, we can perform matched filtering to identify the low-frequency signal in real LIGO data, following \cite{2024PhRvL.133w1401R}. However, even though the ramp-up frequency of the GW memory is well above the low-frequency wall, we train and apply a linear predictive filter to lessen the noise at and below 50 Hz. see Figs. \ref{fig:L1} and \ref{fig:L1_ASD} show a sample of noise from the O3b run at the Livingston site. Here the purple lines show the noise before the LPF is applied and the magenta line shows the noise after the LPF is applied. Beyond glitches in the data, seen at approximately 480 s and 3,700 s, the overall amplitude of the noise is dramatically reduced. Looking to the amplitude spectral density, we can see the amplitude reduction occurring below 50Hz -- precisely at our low-frequency cutoff. This noise reduction technique is not alone in reducing this low-frequency noise. Specifically, \cite{2025Sci...389.1012L} highlights noise reduction in the memory band utilizing a neural network reinforced learning policy driven control design. In this study the proposed method reduced the noise contribution from the current linear controller by one to two orders of magnitude in the 10-30 Hz band, which results in a 10 to 100 times increase in reach for a suit of black hole -- black hole mergers.

\begin{figure}
        \centering
        \begin{subfigure}[t]{0.49\textwidth}
            \centering
            \includegraphics[width=\linewidth]{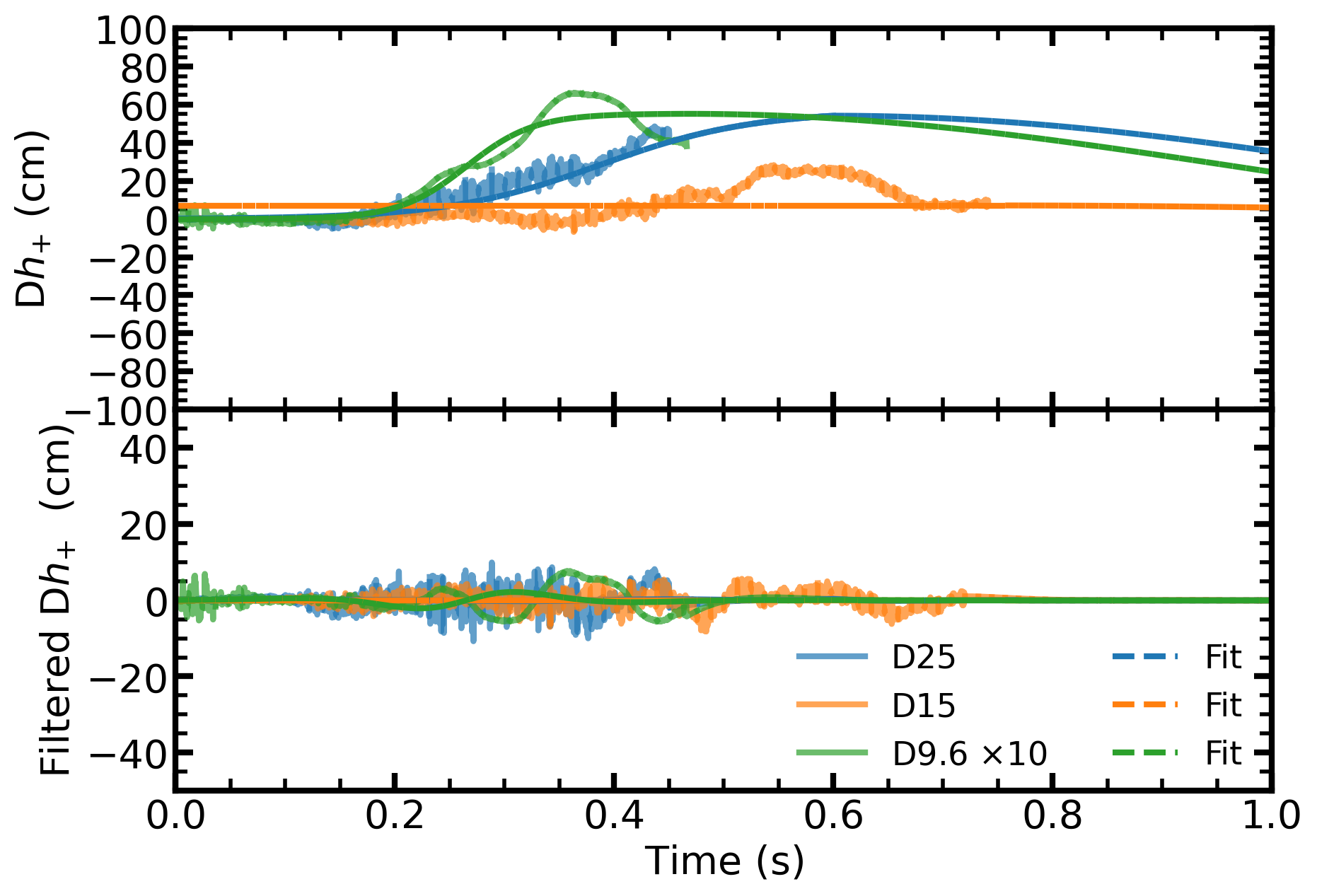} 
            \caption{Gravitational wave strain pre (top) and post (bottom) high-pass filter. The blue lines indicate the signal from D25-3D, the orange lines indicate the signal from D15-3D, and the green line indicates a scaled ($\mathrm{\times 10}$) signal from D9.6-3D. The solid lines represent the data as is from the model and the dashed lines represent the template. } \label{fig:hpf}
        \end{subfigure}
        \begin{subfigure}[t]{0.49\textwidth}
            \centering
            \includegraphics[width=\linewidth]{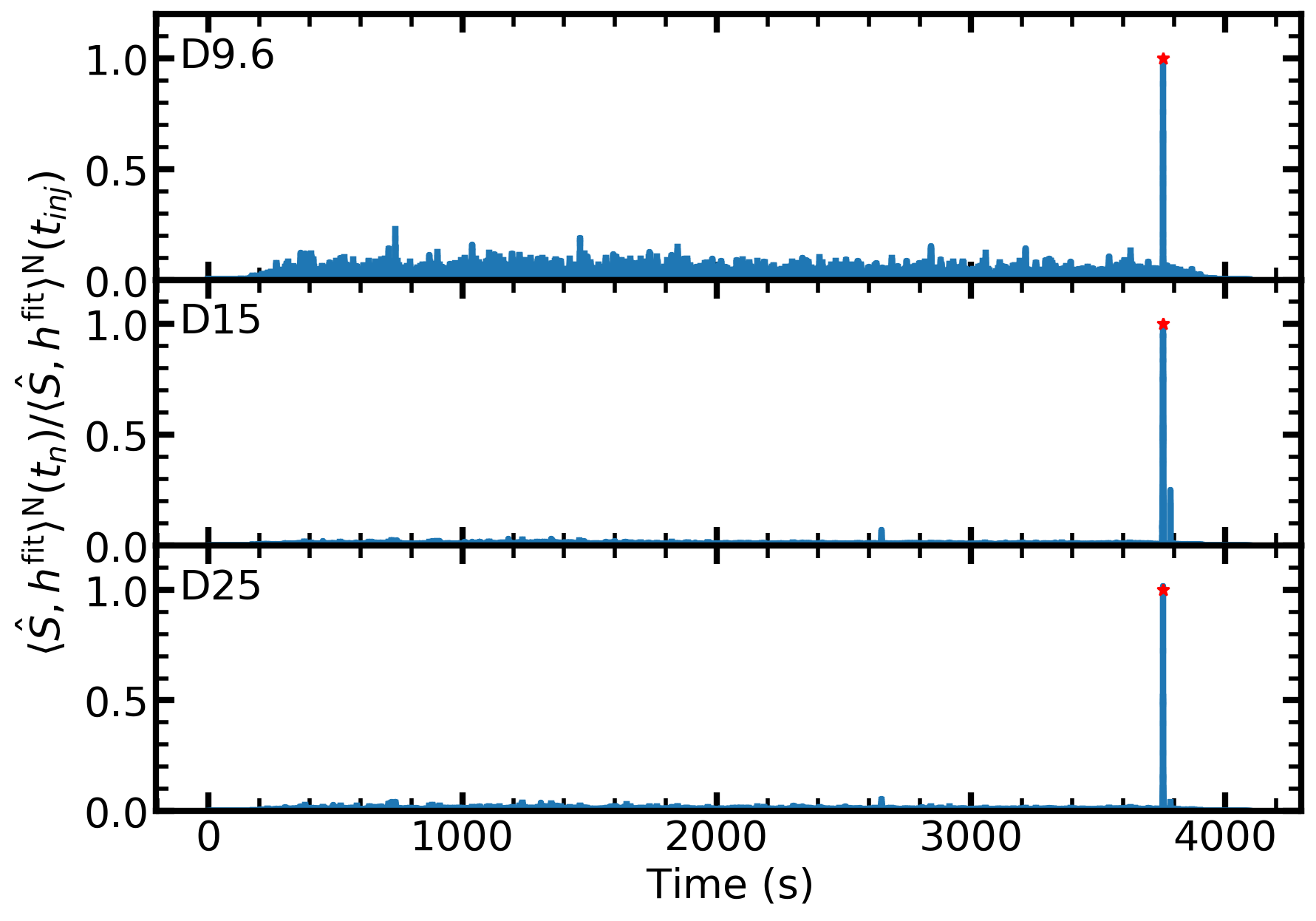} 
            \caption{Normalized correlation of the template search in real LIGO noise. The top, middle, and bottom panel shows the search for an example waveform from the D9.6-3D, D15-3D, and D25-3D models of \cite{2023PhRvD.107d3008M} following \cite{2024PhRvL.133w1401R}.} \label{fig:1kpc_corr}
        \end{subfigure}
        \caption{Detection example in real LIGO data given three example waveforms injected randomly at 1 kiloparsec.}\label{fig:current detection}
    \end{figure}

    By injecting the signals in real LIGO data and then applying the LPF and a high-pass filter, to mimic the filter applied by the Gravitational Wave Open Science Center's publicly available data \cite{2023ApJS..267...29A}, we can then perform a template search to identify the signal in the data (Figs. \ref{fig:current detection}). Here we highlight a different set of signals than those presented in \cite{2024PhRvL.133w1401R}. Fig. \ref{fig:hpf} shows the GW signals and their associated templates before and after the application of the LPF. Fig. \ref{fig:1kpc_corr} shows the final matched-filtering results assuming a signal injected at 1 kpc. Notably, we are clearly able to distinguish the signal from the noise across all three signals. From this we are able to see the robustness of the detection technique and provides even better detection prospects. It is understood that other factors, such as magnetic fields and rotation, progenitor mass, neutron star kicks, and the inclusion of non-standard neutrino physics will alter the GW signature, see \cite{2020MNRAS.492.4613O, 2024ApJ...975...12C, 2019MNRAS.487.1178P, 2022PhRvD.105f3018P, 2025CQGra..42u5002P, 2026PhRvL.136b1201E, 2026arXiv260407878S}. In some cases, the low-frequency portion of the GW signal is larger than those shown herein and may impact the detection horizon utilizing this matched filtering method from the 10 kpc shown in \cite{2024PhRvL.133w1401R} to upwards of 100 kpc in current detectors. 

    Given the sensitivities proposed by the Laser Interferometer Space Antenna, Cosmic Explorer, and the Einstein Telescope, the low-frequency portions of GW signals will become more detectable. Specifically in the case of CE and ET, the overall reduction of the noise and the reduction of the noise between 1-50 Hz will increase our memory detection prospects by upwards of two orders of magnitude. LISA's sensitivity in the nano to millihertz region will not allow us to capture the higher-frequency features of a CCSN GW signal such as the SASI and the oscillations of the proto-neutron star, but will allow us a view of the long-term evolution of the CCSN system. This long-term evolution of CCSNe has recently been highlighted due to the long-time simulations performed by \cite{2013A&A...552A.126W, 2021ApJ...921..113S, 2020ApJ...901..108V}, which show that the asymmetric evolution of the shock wave persists, and thus the gravitational wave memory persists. However, in all of these cases, the treatment of neutrino radiation is reduced for simulations that reach shock break out; therefore, the GW memory sourced from the neutrino field cannot be tracked. From \cite{2024ApJ...975...12C} there is evidence that the neutrino GW memory continues to evolve over multiple seconds.

\begin{figure}
        \centering
        \includegraphics[width=0.5\linewidth]{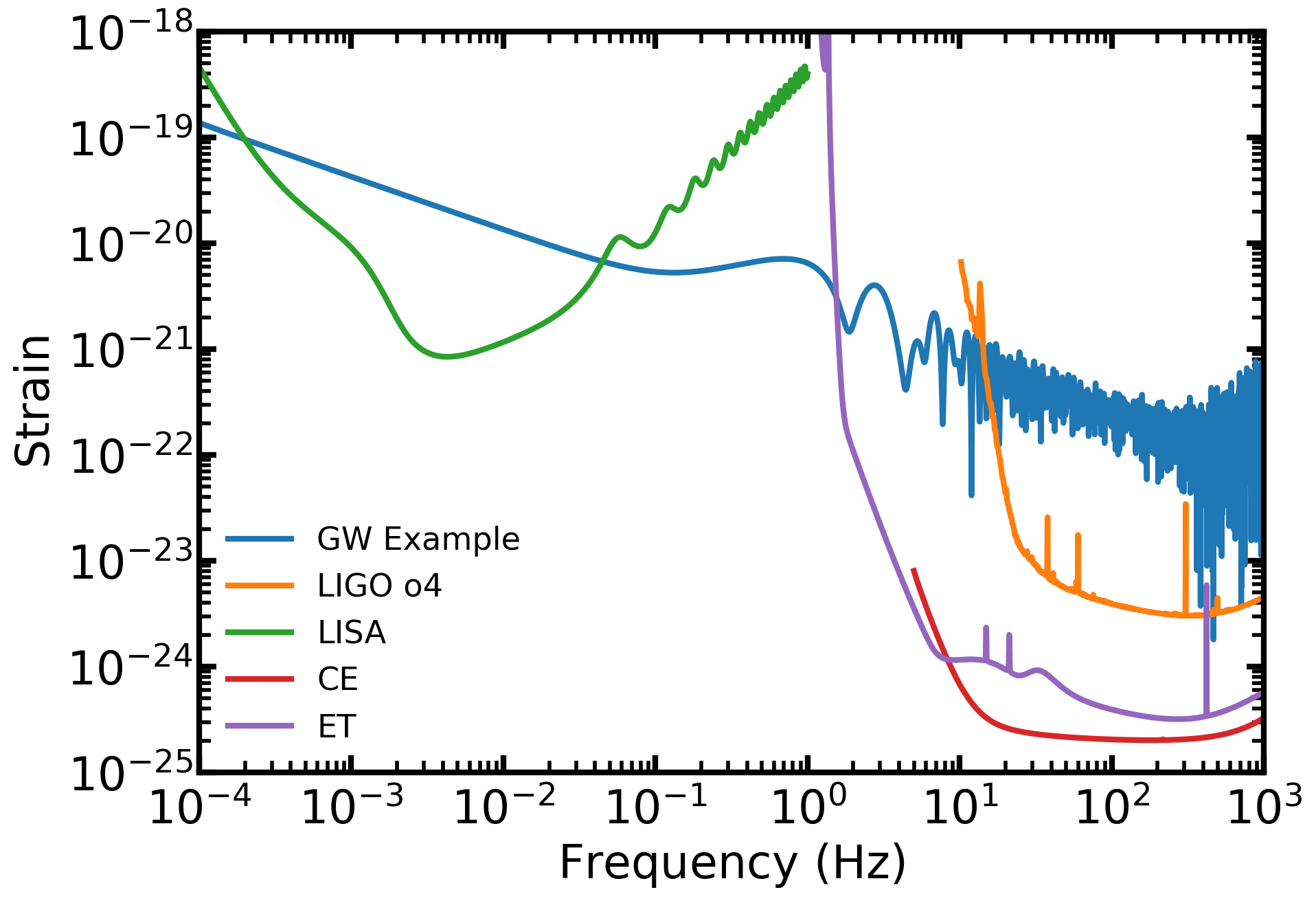}
        \caption{An example GW waveform scaled to a source at 1 kpc, extended to 20000 seconds plotted with the sensitivity curves for current and future detectors. }
        \label{fig:sensitivity}
    \end{figure}

    Assuming the memory from the total GW signal remains roughly constant for 10000 seconds, which is still within shock break out time scale, we can predict the signal interacting with the LISA detector. Fig. \ref{fig:sensitivity} shows the example waveform extended to be a constant value until gradually returning to zero at 10000 seconds. We include this gradual return to zero, or tail, to remove high-frequency ringing induced by the discontinuity in the signal. This tail does not have an impact on the signal at low frequencies, as long as the tail is longer than the inverse of the strain crossing point with the sensitivity curve. For example, given the example waveform, injected at 1 kpc, the crossing frequency for the LIGO sensitivity curve is at 11 Hz, the crossing for ET and extrapolating for CE is around 2 Hz, and the lowest frequency crossing in LISA is at 0.001 Hz. In all of these cases, the tail to remove the high-frequency ringing needs to be longer than 0.1 seconds, 0.5 seconds, and 1000 seconds respectively. For the waveform shown in Fig. \ref{fig:sensitivity} the tail has the form 
    \begin{equation}
        h^{\text{tail}}_{\times/+} = \frac{h^{\text{end}}_{\times/+}}{2} \left[ 1 + \cos(2\pi  f_t(t-t^{\text{end}})) \right],
    \end{equation}
    and for this particular example, $f_{t} = 10^{-5}$ Hz. For these sensitivities, we can estimate the signal-to-noise ratio. Table \ref{tab:snr} shows these SNRs for the sensitivities given. We expect the SNR of CE to be comparable to that of ET, but due to the data for CE stopping at 5 Hz, we lose a lot of the low-frequency information. This highlights the importance of the low-frequency signal in future detectors.
    \begin{table}[]
        \centering
        \begin{tabular}{|c|c|c|}
            \hline
            Detector &Frequency Range & SNR \\
            \hline
            LIGO & $\mathrm{10 \ Hz \ < f < 100 \ Hz}$ & 77.56 \\
            \hline
            LISA & $\mathrm{10^{-5} \ Hz < f < 1 \ Hz}$ & 28.80 \\
            \hline
            CE & $\mathrm{5 \ Hz < f < 100 \ Hz}$ & 3180 \\
            \hline
            ET & $\mathrm{1 \ Hz < f < 100 \ Hz}$ & 1282 \\
            \hline 
            ET & $\mathrm{5 \ Hz < f < 100 \ Hz}$ & 1275 \\
            \hline
        \end{tabular}
        \caption{Signal-to-noise ratio for the example signal extended to 10000 seconds with a $10^{-4}$ Hz tapering injected at 1 kiloparsec.}
        \label{tab:snr}
    \end{table}

    Another opportunity for low-frequency GW detection from CCSNe would utilize the Moon. \cite{2025JCAP...01..108A} propose a new detector based on the Lunar surface that would have a sensitivity between those proposed by LISA and CE/ET. This coverage in the decihertz band would provide complete coverage (when coupled to a LISA/LIGO/ET/CE detection) of the expected frequency range of CCSN GW signatures and explicitly assist in the detection of the linear GW memory from CCSNe. This would also assist in signal reconstruction in current detectors where, coupled with a detection in LIGO, more aggressive noise reduction techniques could be utilized to lower the seismic wall. 
    Recently \cite{2026arXiv260407878S} show how, given a rotating, very massive progenitor and other third-generation space-based detectors, such as the Big Bang Observer (BBO; \cite{2006CQGra..23.4887H}) and the DECi-hertz Interferometer Gravitational Wave Observatory (\cite{2011CQGra..28i4011K}; DECIGO), a low-frequency GW signal could be detected from a source within 100 Mpc. The authors estimate the event rate, given a Salpeter initial mass function (\cite{1955ApJ...121..161S}) and an updated supernova rate from recent All-Sky Automated Survey for Supernovae (ASAS-SN) data (\cite{2025A&A...703A..34P}), to be 2.6 over BBO's 5-year mission. Given these prospective detections and future detectors, the low-frequency region of GWs from CCSNe is becoming a rich and realizable field.

\section{Conclusions and Outlook}
    Given the increased interest in the low-frequency portion of GW signals from CCSNe, a complete treatment of such signals is needed. CCSN models need to include realistic neutrino transport as well as be extended to late times associated with shock break out in order to accurately model the low-frequency predictions of the GW signals. The inclusion of the GWs sourced from the neutrino field increase the amplitude of the low-frequency component of the total signal dramatically, as shown in Figs. \ref{fig:strain} and \ref{fig:fft}, and when models extend a few seconds post bounce the degree of this increase is predicted to be even more dramatic, as in \cite{2024ApJ...975...12C}. In current detectors, evidence of the linear memory is predicted to be detectable given a combination of noise reduction techniques, such as the use of a linear predictive filter or the use of machine learning in the control design at the detectors, and matched filtering techniques similar to those investigated in detail for binary merger searches. The future of low-frequency GWs is bright given the design specifications and sensitivity goals proposed by the Laser Interferometer Space Antenna, Cosmic Explorer, Einstein Telescope, and the Lunar Gravitational-wave Antenna. In all of these cases the evidence of the GW linear memory will be detectable (over different timescales) but when coupled together would provide a more complete picture of the entire CCSN event. Given a detection of this type we could follow the evolution of the shock and predict the global morphology of the explosion and the ejecta; coupled with a neutrino detection we could follow the evolution of the proto-neutron star and make predictions about to kick of the PNS, and given the detection of (or evidence of) the linear gravitational wave memory we could test a yet unconfirmed prediction of general relativity. 

%
% Each of the commands below will create an unnumbered section with the appropriate heading.
% Remove any sections that are not relevant for your article.
% All sections except suppdata will be removed if the [anonymous] option is used.
% See iopjournal-guidelines.pdf for more information.
%

% \ack{}

\funding{ 
A.M. acknowledges support from the National Science Foundation's Gravitational Physics Theory Program through grant PHY-2409148.
H.A. is supported by the Swedish Research Council (Project No. 2020-00452).
M.Z. is supported by the National Science Foundation Gravitational Physics Experimental and Data Analysis Program through awards PHY-2110555 and PHY-2405227.

An award of computer time was provided by the Innovative and Novel Computational Impact on Theory and Experiment (INCITE) program. This research used resources of the Oak Ridge Leadership Computing Facility, which is a DOE Office of Science User Facility supported under contract DE-AC05-00OR22725.

This research has made use of data or software obtained from the Gravitational Wave Open Science Center (gwosc.org), a service of the LIGO Scientific Collaboration, the Virgo Collaboration, and KAGRA. This material is based upon work supported by NSF's LIGO Laboratory which is a major facility fully funded by the National Science Foundation, as well as the Science and Technology Facilities Council (STFC) of the United Kingdom, the Max-Planck-Society (MPS), and the State of Niedersachsen/Germany for support of the construction of Advanced LIGO and construction and operation of the GEO600 detector. Additional support for Advanced LIGO was provided by the Australian Research Council. Virgo is funded, through the European Gravitational Observatory (EGO), by the French Centre National de Recherche Scientifique (CNRS), the Italian Istituto Nazionale di Fisica Nucleare (INFN) and the Dutch Nikhef, with contributions by institutions from Belgium, Germany, Greece, Hungary, Ireland, Japan, Monaco, Poland, Portugal, Spain. KAGRA is supported by Ministry of Education, Culture, Sports, Science and Technology (MEXT), Japan Society for the Promotion of Science (JSPS) in Japan; National Research Foundation (NRF) and Ministry of Science and ICT (MSIT) in Korea; Academia Sinica (AS) and National Science and Technology Council (NSTC) in Taiwan.
}
% This section is a list of funder names and grant numbers
\data{The data presented here is part of the \textsc{Chimera} group's GW data release which can be found at https://doi.ccs.ornl.gov/dataset/847fc720-6ff7-50eb-a747-12fbb23038db .}
\printbibliography
\end{document}